\documentclass[varenna]{cimento}
\usepackage{graphicx} 
\usepackage{amsmath} 
\usepackage{amssymb} 
\usepackage{bm}
\usepackage{xcolor}

\title{Multicomponent spin mixtures of two-electron fermions}

\author{Leonardo Fallani}

\institute{Dipartimento di Fisica e Astronomia,
Università degli Studi di Firenze (Italy)\\LENS European Laboratory for Nonlinear Spectroscopy (Italy)\\CNR-INO Istituto Nazionale di Ottica del CNR, Sezione di Sesto Fiorentino (Italy)}

\begin{document}

\maketitle

\begin{abstract}
These lecture notes contain an introduction to the physics of quantum mixtures of ultracold atoms trapped in multiple internal states. I will discuss the case of fermionic isotopes of alkaline-earth atoms, which feature an intrinsic SU($N$) interaction symmetry and convenient methods for the optical manipulation of their nuclear spin. Some research directions will be presented, with focus on experiments performed in Florence with nuclear-spin mixtures of $^{173}$Yb atoms in optical lattices.
\end{abstract}

\tableofcontents


\section{Introduction}

Two-level systems are ubiquitous in quantum physics. An example is the spin $S=1/2$ of elementary particles, which leads to the quantization of the spin projection quantum number in two possible states $|$$\uparrow \rangle$ ($m_S=+1/2$) and $|$$ \downarrow \rangle$ ($m_S=-1/2$). From the existence of the spin degree of freedom many fundamental effects emerge in condensed matter physics, among these the magnetic and superconducting properties of materials. Some of these effects have been successfully explored in quantum simulation experiments with binary spin mixtures of ultracold atoms, as discussed during the School and in several chapters of these Proceedings. Needless to say, two-level quantum systems are also a central paradigm in quantum information science, where they are treated as qubits, i.e., quantum bits of information, which can be encoded, processed and detected thanks to their coherent interaction with tailored external fields.

Multicomponent quantum systems, i.e., quantum systems featuring higher-dimensional internal Hilbert spaces, significantly expand the realm of physical effects and applications. Interacting multicomponent systems can exhibit different quantum phases from those of their two-component counterparts, and they can allow for new strategies for the manipulation of quantum information encoded in higher-dimensional quantum bits, i.e., \textit{qudits}. Multicomponent systems can be realized when particles with spin higher than $S=1/2$ are considered, or when more than one quantum degree of freedom is involved. Both those scenarios can naturally emerge from the internal structure of atoms, thanks to the large nuclear spin of some isotopes and/or from the multiple quantum numbers labelling the atomic states, provided that those degrees of freedom can be efficiently controlled and probed with external fields, preserving quantum coherence.

In these lecture notes I will focus on some experimental possibilities that are disclosed by fermionic isotopes of two-electron atoms, such as $^{173}$Yb or $^{87}$Sr, which can form multicomponent quantum spin mixtures with a controllable number of components. These isotopes are characterized by a zero electronic angular momentum in their lowest-energy electronic state ($J=0$) and a nonzero nuclear spin ($I=5/2$ for $^{173}$Yb and $I=9/2$ for $^{87}$Sr), which is decoupled from the electronic degree of freedom. The nuclear-spin manifold offers a large-sized internal Hilbert space (with dimension $N=2I+1=6$ for $^{173}$Yb and $N=2I+1=10$ for $^{87}$Sr), which can be manipulated with high accuracy taking advantage of optical transitions towards excited states, where the hyperfine interaction couples the nuclear spin with the electronic degree of freedom.  Section \ref{sec:su(n)} of this chapter is devoted to an introduction to the SU($N$) interaction symmetry that naturally emerges in these systems, while the experimental techniques for their optical manipulation will be described in Section \ref{sec:exptechniques}. 

These atoms offer exciting perspectives for the quantum simulation of multi-component quantum systems, both for fundamental studies and for applications. Large-$N$ quantum models were  introduced decades ago in the context of theoretical physics, often as mathematical tools to treat strongly correlated quantum systems, but now these models can be realized experimentally and their interesting quantum properties probed directly, as it will be described in Section \ref{sec:exp-su(n)lowdim}. They also allow for the realization of multicomponent lattice systems with interesting connections with the physics of multiorbital materials, the electronic properties of which cannot be simply described in terms of single-band spin-1/2 models, as described in Section \ref{sec:exp-su(n)-lattices}. The possibility of coherent manipulation of the spin manifold also allows for the implementation of new concepts for quantum simulation, based on the realization of so-called \textit{synthetic dimensions}, which provide new avenues for the study of artificial gauge fields and quantum systems with nontrivial topology, as discussed in section \ref{sec:exp-synthetic}. 

Two-electron atoms also offer the possibility of producing quantum mixtures of atoms in different electronic states, profiting of the ultranarrow optical transitions exploited in optical clock experiments. Although the focus of these lectures is on multicomponent quantum mixtures in the electronic ground state, an introduction to the possibilities opened by the excitation of the electronic degree of freedom is presented in the concluding section \ref{sec:furtherdirections}.


\section{Interactions in two-electron fermions and SU($N$) physics}
\label{sec:su(n)}

We start by considering the properties of ultracold collisions between two fermionic two-electron atoms. The interaction potential $V(R)$ governing the collision is a short-ranged potential, characterized by a $\sim R^{-6}$ dependence at large interatomic distance $R$, describing the van der Waals attraction between mutually induced electric dipole moments. Magnetic dipole-dipole interactions are irrelevant for this class of atoms, being them suppressed by the purely nuclear character of their spin\footnote{Nuclear magnetic moments are proportional to the nuclear magneton $\mu_N=e\hbar/2m_p$, that is smaller than the Bohr magneton $\mu_B=e\hbar/2m_e$ by a factor corresponding to the proton-to-electron mass ratio $m_p/m_e\simeq 1836$.}. 

When two atoms with angular momenta $\mathbf{F}_1$ and $\mathbf{F}_2$ collide at ultralow temperatures under the action of the short-ranged potential $V(R)$, only $s-$wave collisions are energetically allowed and the scattering properties only depend on a set of scalar quantities that take the name of {\it scattering lengths} (see Ref. \cite{inguscio1999} for seminal lecture notes on the physics of ultracold collisions). Let's indicate the total angular momentum of the atom pair with $\boldsymbol{\mathcal{F}}=\mathbf{F}_1+\mathbf{F}_2$, which is quantized in integer steps in the range $\mathcal{F}=|F_1-F_2|, \ldots, F_1+F_2$. It is very common to simplify the scattering problem by introducing a zero-ranged effective pseudo-potential, which can be written as
\begin{equation}
V_{\mathbf{eff}}(R)=\sum_{\mathcal{F}}{g_\mathcal{F} P_\mathcal{F} \delta(R)} \; \mathrm{,}
\end{equation}
where $\delta(\cdot)$ is the Dirac delta function, $P_\mathcal{F} = \sum_{\mathcal{M}=-\mathcal{F}}^{+\mathcal{F}}{|\mathcal{F}, \mathcal{M}\rangle \langle \mathcal{F}, \mathcal{M}|}$ is the projection operator on the manifold corresponding to total angular momentum $\mathcal{F}$, $\mathcal{M}$ is the angular momentum projection quantum number and $g_\mathcal{F}$ are the interaction coupling constants
\begin{equation}
g_\mathcal{F}=\frac{4\pi\hbar^2 a_\mathcal{F}}{M} \, ,
\end{equation}
where $a_\mathcal{F}$ is the s-wave scattering length for the $\mathcal{F}$ scattering channel, $\hbar$ is the reduced Planck constant and $M$ is the atomic mass.

The scattering lengths $a_\mathcal{F}$ depend on the characteristics of the short-ranged potential $V(R)$, which arises from the modification of the electronic orbitals when two atoms approach each other (in the spirit of the Born-Oppenheimer principle in molecular physics). Generally speaking, the electronic orbitals  depend on the angular momentum state because of the hyperfine coupling, and so the scattering lengths exhibit a dependence on $\mathcal{F}$. In fermionic isotopes of two-electron atoms, however, the angular momentum $\mathbf{\mathcal{F}}$ has a purely {\it nuclear} character and there is no hyperfine interaction: since the nuclear spin is decoupled from the electronic degree of freedom, $V(R)$ does not depend on $\mathcal{F}$ and so the scattering lengths don't.

Several theoretical works (see Ref. \cite{cazalilla2014} and references therein) have highlighted that this property of fermionic isotopes of two-electron atoms can be described in terms of an exact emergent SU($N$) symmetry. This symmetry originates from the independence of the interaction properties from the specific nuclear-spin states occupied by the atoms. On more formal grounds, the many-body Hamiltonian for an interacting nuclear-spin mixture of two-electron fermions can be written as:
\begin{equation}
\hat{H}=\int d\mathbf{r} \left(
\frac{\hbar^2}{2M} \sum_m \nabla\Psi_m^\dagger \nabla\Psi_m^{ } + \frac{g}{2} \sum_{m,n\neq m} \Psi_m^\dagger \Psi_n^\dagger \Psi_n^{ } \Psi_m^{ }
\right) \, ,
\label{eq:manybody-su(n)}
\end{equation}
where $\Psi_m$ is the field operator for fermions in nuclear-spin projection state $m \in \left\{ -I, \ldots, +I \right\}$ and now the interaction coupling constant $g$ does not depend on the angular-momentum state anymore. It is easy to show that this Hamiltonian commutes with any spin-permutation operator $\hat{S}_{mn}$ that changes the spin index from $n$ to $m$
\begin{align}
\label{eq:SUN-Snm}
&\hat{S}_{mn} = \int  d\mathbf{r} \Psi_m^\dagger \Psi_n^{ } \\ 
&\left[\hat{H},\hat{S}_{mn}\right] = 0 \, ,
\label{eq:SUN-Hcommutes}
\end{align}
and it can be verified that these operators satisfy commutation relations
\begin{equation}
\left[ \hat{S}_{mn}, \hat{S}_{pq} \right] = \delta_{mq}\hat{S}_{pn} - \delta_{pn}\hat{S}_{mq} \, ,
\end{equation}
which have the mathematical structure exhibited by the generators of an SU($N$) symmetry group, with $N=2I+1$. 

In addition to highlighting the SU($N$) symmetry of the Hamiltonian, the equations above have another important implication. When $m=n$ the operator $\hat{S}_{mn}$ reduces to the number operator for fermions in state $m$, whose expectation value
\begin{equation}
N_m = \langle \hat{S}_{mm} \rangle
\end{equation}
is conserved by the action of $\hat{H}$ according to Eq.  (\ref{eq:SUN-Hcommutes}), i.e., the number of atoms in each $m$ is constant. We note that this is not generally true for ultracold quantum mixtures: in particular, in alkali atoms the dependence of $g_\mathcal{F}$ on $\mathcal{F}$ leads to two-body spin-exchange interactions, which induce a dynamics in the spin sector, i.e., $N_m$ are not constant. Other contributions in this volume describe the scenarios that are opened by these processes, e.g. in the context of spinor Bose-Einstein condensates.

In order to make the discussion above more concrete, let's consider the two simplest cases: SU(2) and SU(3). For the SU(2) group the generators can be recast in terms of the well-know 2x2 Pauli matrices describing generic rotations of a spin 1/2 
\begin{equation}
\sigma_x = \left(\begin{matrix}0&1\\1&0\end{matrix}\right)\,,\;
\sigma_y = \left(\begin{matrix}0&-i\\i&0\end{matrix}\right)\,,\;
\sigma_z = \left(\begin{matrix}1&0\\0&-1\end{matrix}\right)\;,
\end{equation}
with the addition of the identity operator. For the SU(3) group the generators are the 3x3 Gell-Mann matrices, which generalize the action of the Pauli matrices to a three-dimensional Hilbert space:
\begin{multline}
\lambda_1 = \left(\begin{matrix}0&1&0\\1&0&0\\0&0&0\end{matrix}\right)\,,\;
\lambda_2 = \left(\begin{matrix}0&-i&0\\i&0&0\\0&0&0\end{matrix}\right)\,,\;
\lambda_3 = \left(\begin{matrix}1&0&0\\0&-1&0\\0&0&0\end{matrix}\right)\,, \\
\lambda_4 = \left(\begin{matrix}0&0&1\\0&0&0\\1&0&0\end{matrix}\right)\,,\;
\lambda_5 = \left(\begin{matrix}0&0&-i\\0&0&0\\i&0&0\end{matrix}\right)\,,\;
\lambda_6 = \left(\begin{matrix}0&0&0\\0&0&1\\0&1&0\end{matrix}\right)\,, \\
\lambda_7 = \left(\begin{matrix}0&0&0\\0&0&-i\\0&i&0\end{matrix}\right)\,,\;
\lambda_8 = \frac{1}{\sqrt{3}}\left(\begin{matrix}1&0&0\\0&1&0\\0&0&-2\end{matrix}\right)\,.
\end{multline}
The Gell-Mann matrices find an important application in quantum chromodynamics, as they are used to describe the SU(3) color symmetry of the strong interaction between quarks.

Because of this connection with concepts and topics of high-energy physics, in the context of ultracold SU($N$) quantum mixtures the internal degree of freedom that comes from the spin can also be dubbed as a \textit{color}, or \textit{flavour} degree of freedom. This comes in analogy with the color symmetry and flavour symmetry that appear in the physics of strong interactions, and also from the consideration that the spin orientation does not play any role in the physics of the problem, other than marking the distinguishability of the particles.  This analogy with symmetries in high-energy physics should be elaborated with care, as there are important differences between the SU($N$) symmetry that we have described in this section and e.g. the SU(3) color symmetry of quantum chromodynamics, the former being a \textit{global} symmetry, the latter being a \textit{local} symmetry of the underlying gauge theory.

For more details on the emergence and consequences of the SU($N$) symmetry in ultracold quantum mixtures the reader can refer to specialized reviews \cite{cazalilla2014}. In the next sections of these lecture notes we will focus on some specific examples, connected with experimental work performed at LENS \& University of Florence with nuclear-spin mixtures of fermionic $^{173}$Yb atoms.


\section{Experimental techniques}
\label{sec:exptechniques}

\subsection{Spin manipulation and detection}

In this section we will present the most important techniques for the experimental control and detection of the nuclear spin degree of freedom, with direct reference to the case of $^{173}$Yb atoms, which feature a nuclear-spin $I=5/2$, thus 6 spin projection components $m \in \left\{ -5/2 ,-3/2, \ldots ,+5/2 \right\}$.

A direct control of nuclear spins by static and/or time-dependent magnetic fields would be rather inconvenient for these atoms. As a matter of fact, the magnetic dipole moment $\bm{\mu}=g_I \mu_N \mathbf{I}$ of a nuclear spin $\mathbf{I}$ is much smaller than the magnetic dipole moment $\bm{\mu} = g_J \mu_B \mathbf{J}$ associated to the electronic angular momentum $\mathbf{J}$, as the first is proportional to the nuclear magneton $\mu_N=e\hbar/2m_p$ while the second is proportional to the Bohr magneton $\mu_B=e\hbar/2m_e \simeq 1836 \mu_N$ (where $m_e$ and $m_p$ are the electron and proton masses, respectively, and $g_I$ and $g_J$ are Land\'e factors on the order of one). For $^{173}$Yb atoms in their electronic ground state $^1S_0$ the Zeeman shift  $\Delta E = \langle -  \bm{\mu} \cdot \mathbf{B} \rangle$ induced by a magnetic field $\mathbf{B}$ amounts to just $\Delta E / h \sim 200$ Hz/G $\times$ $mB$. This weak magnetic sensitivity of nuclear spins\footnote{It is well known that very large magnetic fields (on the order of 1 T or more) have to be used in NMR setups to obtain resolved spectroscopic signals or high-resolution NMR imaging from the tiny magnetic dipole moments of nuclear spins.} makes most of the experimental techniques usually employed for alkali atoms, such as magnetic trapping or magnetic Stern-Gerlach detection, not practical. 

Thus, the experimental methods for the spin manipulation of two-electron fermions are all based on optical techniques, more specifically on the interaction of the atoms with polarized light. This can be realized by coupling the atoms in the electronic ground state $^1S_0$ to a state that is magnetically sensitive such as the $^3P_1$. In $^{173}$Yb the   $^1S_0$ $\rightarrow$ $^3P_1$ transition has a wavelength $\lambda=556$ nm and a natural linewidth $\gamma/2\pi \simeq 180$ kHz. In the $^3P_1$ state the nuclear spin is coupled to the electronic angular momentum via the hyperfine interaction, leading to a $^3P_1$ hyperfine triplet ($F=3/2$, 5/2, 7/2) with total separation $\Delta/2\pi  \simeq 6.2$ GHz. Different coupling configurations can be implemented, leading to different manipulation and detection techniques.
\hfill\break

\begin{figure}[p]
\begin{center}
\includegraphics[height=15.1cm]{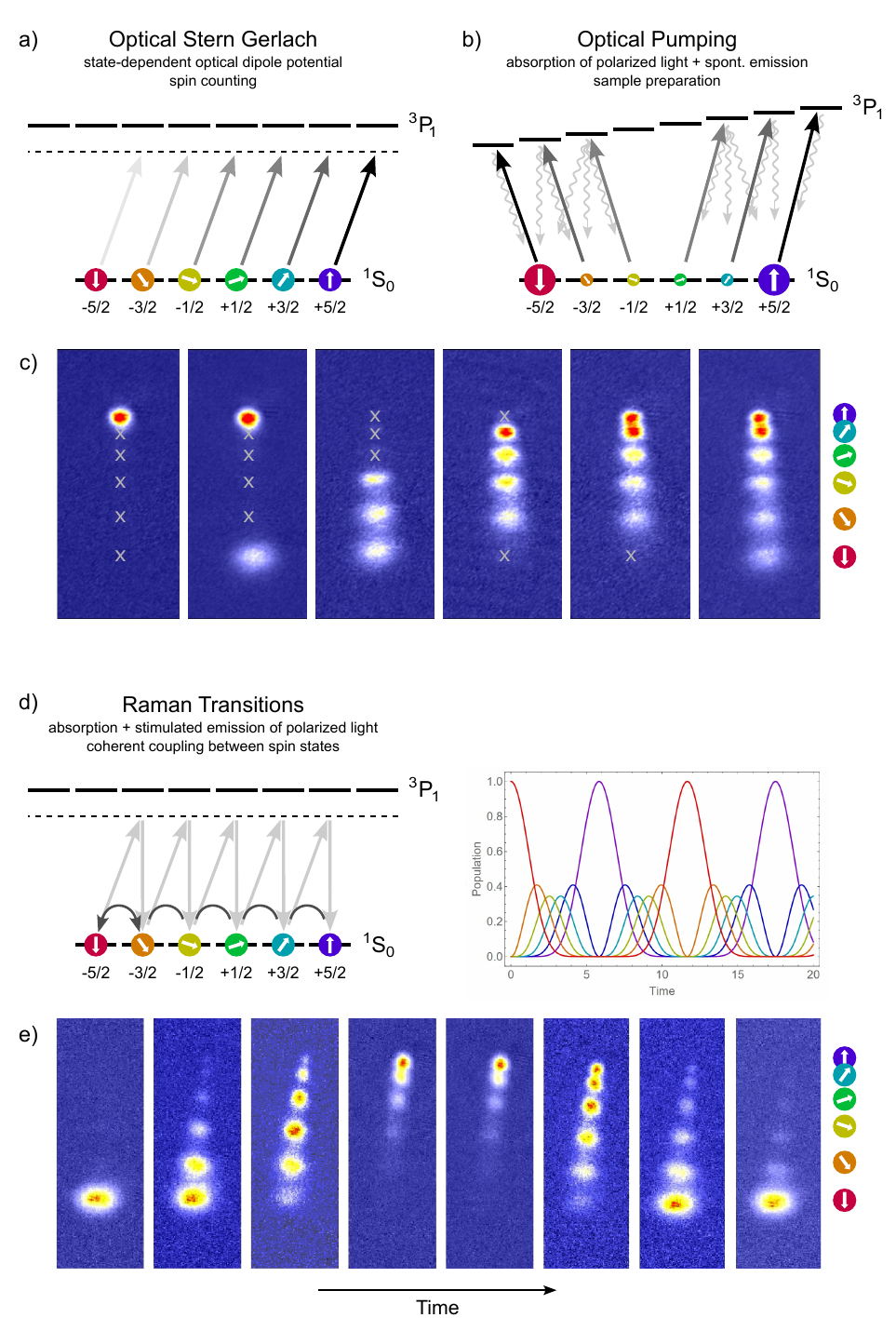}
\end{center}
\caption{Optical manipulation of nuclear-spin mixtures of two-electron $^{173}$Yb fermions: a) in the Optical Stern Gerlach (OSG) detection a spin-dependent coupling to state $^3P_1$ induces a spin-dependent optical dipole force; b) absorption of polarized light and spontaneous emission is used to control the population within the nuclear-spin manifold; c) experimental OSG images for different optically-pumped mixtures; d) coherent spin-flip couplings can be induced in a two-photon Raman process driven by two laser beams with different polarizations: this realizes a coherent evolution in the nuclear-spin manifold, as shown by the Rabi dynamics in the right panel; e) experimental OSG images of the dynamics of a spin-polarized sample driven by the Raman coupling. The experimental images are from the $^{173}$Yb experiment in Florence.} \label{fig:manipulation}
\end{figure}

\textit{Optical Stern-Gerlach detection.}
In order to probe the nuclear-spin composition of the fermionic mixture it is convenient to image the atomic cloud after a Stern-Gerlach deflection, which is obtained via the optical dipole force induced by a gradient of light intensity. We recall that an atom illuminated by a non-uniform, non-resonant light field is subjected to a spatially varying optical dipole potential of the form
\begin{equation}
V_{\mathrm{dip}}(\mathbf{r}) \propto \alpha I(\mathbf{r}) \, ,
\label{eq:vdip}
\end{equation}
where $\alpha$ is the atomic polarizability, depending on the atomic internal structure and on the frequency of the light \cite{grimm2000}. For moderate detunings $\delta / 2\pi \approx \times 1$ GHz -- much larger than the transition linewidth $\gamma$ in order to suppress absorption, but not much larger than the hyperfine structure $\Delta$ -- the polarizability $\alpha$ strongly depends\footnote{This dependence can be expressed in terms of the Clebsch-Gordan coefficients entering the description of light-matter interaction in multilevel atoms. Far from saturation, $\alpha$ is proportional to the coefficient $\langle F,m ; 1,\sigma | F',m+\sigma \rangle$, where $F$ and $F'$ are the initial and final angular momentum states and $\sigma$ describes the polarization state of light (0 for linear polarization, $\pm 1$ for circular polarization). For more details see e.g. Ref. \cite{steck2022}.} on $m$, leading to a spin-dependent optical dipole potential $V_{\mathrm{dip}}^m(\mathbf{r})$, thus to a spin-dependent optical force that deflects the atoms according to their spin state (this is exemplified by the intensity of the arrows in Fig. \ref{fig:manipulation}a, describing the variation of the optical coupling with $m$). After this optical Stern-Gerlach (OSG) deflection, the atoms are detected with standard time-of-flight absorption imaging, which allows for the determination of the atom number in each spin state, as shown in Fig. \ref{fig:manipulation}c.
\hfill\break

\textit{Optical pumping.}
Different spin mixtures can be prepared by relying on different preparation methods involving polarized light. The simplest scheme is incoherent optical pumping (OP), which can be performed by exploiting the spontaneous emission following a resonant excitation of the atoms on the $^1S_0$ $\rightarrow$ $^3P_1$ transition (see Fig. \ref{fig:manipulation}b). By exploiting the Zeeman shift of the excited states and engineering appropriate sequences of OP pulses, it is possible to create arbitrary spin mixtures, as exemplified in the images in Fig. \ref{fig:manipulation}c. This technique is useful for the study of fermionic mixtures with different SU($N$) symmetries, according to the number of populated states, as in the experiment discussed in Sec. \ref{sec:exp-su(n)lowdim}.
\hfill\break

\textit{Two-photon Raman coupling.}
In addition to incoherent optical pumping, the nuclear spin can be controlled in a {\it quantum coherent} way by taking advantage of two-photon Raman transitions. The atoms are illuminated by two laser beams with two different polarization states and frequencies $\omega_1$ and $\omega_2$ respectively. If the energy difference between the photons $\hbar(\omega_2-\omega_1)$ is resonant with the energy separation between different states in the $^1S_0$ nuclear-spin manifold, a two-photon transition can take place with the absorption of a photon from the first beam and the stimulated emission of a photon in the mode of the second beam. In order to suppress single-photon absorption both the photons are detuned from the $^1S_0$ $\rightarrow$ $^3P_1$ transition by a detuning $\delta$ (typically in the GHz range) much larger than the natural linewidth of the transition $\gamma$. In this regime the Raman process can be described as an effective coherent coupling between the spin states with a two-photon Rabi frequency
\begin{equation}
    \Omega = \frac{\Omega_1^* \Omega_2} {2 \delta} \; ,
\label{eq:rabiraman}
\end{equation}
where $\Omega_1$ and $\Omega_2$ are the Rabi frequencies of the single-photon excitations \cite{steck2022}. This coupling can be extended to more than two spin states, as shown in Fig. \ref{fig:manipulation}d-e, which display Rabi dynamics in the 6-dimensional nuclear spin manifold of $^{173}$Yb atoms, driven by two Raman beams with linear and circular polarization respectively, thus inducing two-photon transitions with $\Delta m=\pm 1$. The population oscillating forth and back in the nuclear-spin manifold is clearly reminiscent of Rabi flopping in a 2-level system.
\hfill\break

\textit{Spin-selective imaging.}
Finally, we mention the possibility to perform state-selective imaging, where the properties of each component of the nuclear-spin mixture can be detected individually. Rather than relying on the OSG deflection, which is prone to induce artifacts in the cloud shape (because of residual curvatures in the light intensity gradient),  spin-selective imaging can be performed by relying on the OP approach presented before. Before the imaging pulse, an engineered sequence of resonant OP pulses is triggered, in such a way as to ``blast" the mixture components in unwanted spin states, in order to keep only the chosen spin state for imaging (as in the experiments discussed in Sec. \ref{sec:exp-synthetic-currents}). Again, this can be performed conveniently on the $^1S_0$ $\rightarrow$ $^3P_1$ transition, which is characterized by a small linewidth $\gamma$ and by a substantial Zeeman shift in the excited state, enabling for an accurate spin-selective excitation.

\subsection{Optical lattices}

In the following sections we will focus on the physics of multicomponent nuclear-spin mixtures trapped in reduced dimensionality and/or in lattice structures. These configurations can be realized experimentally by proper arrangements of \textit{optical lattices}.

\begin{figure}[t!]
\begin{center}
\includegraphics[width=0.9\columnwidth]{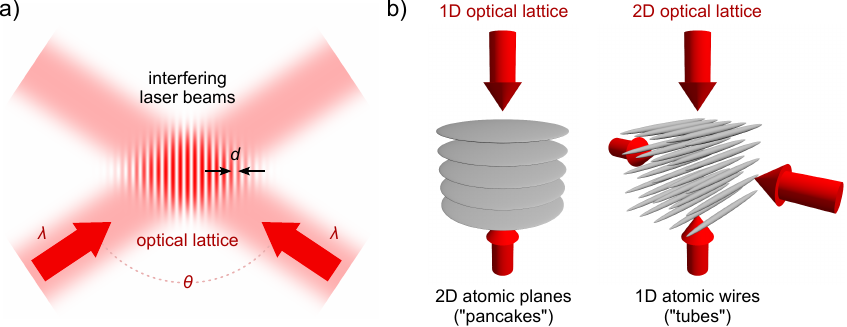}
\end{center}
\caption{a) Scheme of the standing-wave intensity pattern resulting from the interference of two intersecting laser beams; b) schemes of deep optical lattices creating arrays of low-dimensional quantum gases, where the atomic motion is restricted either to 2D or 1D.} \label{fig:lattice}
\end{figure}

An optical lattice is the optical dipole potential that is created by 
a periodic modulation of light, such as that produced by the standing-wave interference of two laser beams with the same frequency, same polarization and different direction of propagation. If two beams cross at an angle $\theta$, as sketched in
Fig. \ref{fig:lattice}a, the distance between two adjacent
maxima (or minima) of the resulting interference pattern is
\begin{equation}
d= \frac{\lambda}{2 \sin{(\theta/2)}} \; .
\end{equation}
The simplest experimental setting is provided by a single laser beam propagating along $\hat{x}$ that interferes with its retroreflection propagating along $-\hat{x}$: in this counterpropagating configuration the  standing-wave pattern has an
intensity modulation of period $d=\lambda/2$ and the resulting optical dipole potential (already introduced in Eq. (\ref{eq:vdip})) is
\begin{equation}
V_\mathrm{dip}(x) = V_0 \cos^2(kx) \; , \label{eq:optlatt}
\end{equation}
where $k=2\pi / \lambda$ is the laser wavenumber and $V_0$ is the depth of the periodic potential, proportional to the maximum intensity of the standing-wave pattern. $V_0$ is often measured in units of the recoil energy $E_R=\hbar^2 k^2
/ 2M= h^2 / 8 M d^2$ (where $M$ is the atomic mass), which physically corresponds to the
kinetic energy an atom at rest acquires after absorption of one lattice photon. When $V_0<0$ the atoms are trapped around the maxima of intensity, which define the positions of the lattice sites.

When the depth of the optical lattice is sufficiently strong (typically $|V_0| \gtrsim 20 E_R$) quantum tunnelling between different sites of the optical lattice can be neglected, i.e., each lattice site provides an independent trap. Furthermore, at the ultralow temperatures of quantum degenerate gases, the atomic motion is typically restricted to the ground state of the individual traps. This makes optical lattices a very convenient tool to realize low-dimensional quantum gases, as pictured in Fig. \ref{fig:lattice}: a deep 1D optical lattice can be used to create a stack of 2D quantum gases, while a 2D lattice can realize an array of 1D atomic quantum wires (as further discussed in Sec. \ref{sec:exp-su(n)lowdim}).

For lower depths of the optical lattice quantum tunnelling between the lattice sites becomes important. In the tight-binding limit (typically reached for $|V_0| \gtrsim 4 E_R$) only tunnelling between nearest-neighboring sites is considered and the atomic system provides a perfect realization of a tight-binding model
\begin{equation}
\hat{H} = -t \sum_{\langle i,j \rangle} \hat{c}_i^\dagger \hat{c}_j^{ }  \, , \label{eq:tightbinding}  
\end{equation}
where $\hat{c}_j$ is the destruction operator of a particle in site $j$ and $t$ (also called {\it hopping energy} or {\it tunnelling energy}) is the quantum amplitude for a tunnelling process of a particle between nearest-neighboring sites $j$ and $i$. In the tight-binding limit $t$ depends quasi exponentially on the lattice depth according to the relation \cite{zwerger2003}
\begin{equation}
t \simeq \frac{4}{\sqrt{\pi}} E_R s^{3/4} e^{-2\sqrt{s}} \, ,
\end{equation}
where $s=|V_0|/E_R$ measures the lattice depth in units of the recoil energy. 

From the tight-binding model of Eq. (\ref{eq:tightbinding}) all the core concepts of solid-state physics, starting from the quantum description of the particle motion in terms of Bloch wavefunctions and energy bands, follow. Optical lattices are, indeed, one of the most powerful tools for quantum simulation, as they grant the possibility to synthesize artificial materials with full control on the microscopic Hamiltonian and parameters tunability. In the next sections I will discuss those relevant facts and features of optical lattices that are most directly connected with the main topic of my lectures. For an in-depth introduction to the physics of ultracold atoms in optical lattices the reader can refer to specialized books \cite{lewenstein2012,inguscio2013}, topical reviews \cite{bloch2005,morsch2006, bloch2008, gross2017, schaefer2020} and comprehensive chapters in the proceedings of previous Enrico Fermi schools \cite{inguscio1999,inguscio2007,inguscio2016}.


\section{Experiments with interacting SU($N$) mixtures in one dimension}
\label{sec:exp-su(n)lowdim}

The experimental realization of low-dimensional ultracold gases offers important opportunities in the context of quantum physics. As a matter of fact, there are a number of quantum phenomena that strongly depend on the dimensionality $d$ of the system. An example is given by quantum phase transitions, that can or cannot occur depending on the system dimensionality, with critical exponents also depending on $d$. Low-dimensional quantum systems are particularly interesting when  atom-atom interactions are considered, as interaction-induced correlations become stronger and stronger as the dimensionality of the system is reduced. Notable examples of low-dimensional ultracold physics include the study of the Berezinski-Kosterlitz-Thouless transition for a 2D interacting Bose gas \cite{hadzibabic2006} or the investigation of Luttinger physics and the demonstration of Tonks-Girardeau gases in 1D systems \cite{kinoshita2004,paredes2004}. Quantum physics of interacting 1D systems is especially interesting because there the effect of quantum correlations is maximally strong, due to the impossibility of particles to swap positions without avoiding each other (see Ref. \cite{giamarchi2004} for a comprehensive review of this topic).

In this section I will present the results of experiments performed in Florence with arrays of 1D SU($N$) $^{173}$Yb gases \cite{pagano2014}, prepared by using a strong 2D optical lattice as discussed in the previous section. The experimental platform allows for the realization of the multicomponent version of the Gaudin-Yang model
\begin{equation}
\hat{H}=\sum_m \int dx \, \Psi_m^\dagger \left(-
\frac{\hbar^2}{2M} \frac{d^2}{dx^2} + V(x) \right) \Psi_m^{ }  + \frac{g_{\mathrm{1D}}}{2} \sum_{m,n\neq m} \int dx \, \Psi_m^\dagger \Psi_n^\dagger \Psi_n^{ } \Psi_m^{ }  \, ,
\label{eq:gaudinyang-su(n)}
\end{equation}
which is the 1D version of Eq. (\ref{eq:manybody-su(n)}), with a modified interaction constant $g_{\mathrm{1D}}$ that takes into account the dimensional reduction \footnote{
Far from confinement-induced resonances -- which is the relevant case for the experiments discussed in the following -- the 1D interaction constant can be written as $g_{\mathrm{1D}} = 4\hbar^2 a / (M a_\perp^2)$, where $a$ is the 3D scattering length and $a_\perp$ is the harmonic oscillator length associated to the transverse confinement in the 2D lattice sites  \cite{olshanii1998}.} \cite{olshanii1998} and the addition of a state-independent external potential $V(x)$, typically a harmonic trap $V(x)= m \omega^2 x^2/2$ with angular frequency $\omega$. This model (see Ref. \cite{guan2013} for a comprehensive review) can be considered as the fermionic counterpart of the Lieb-Liniger model describing the physics of a 1D interacting bosonic system. 

The experimental results discussed below refer to balanced $^{173}$Yb spin mixtures with different number of components $N$, obtained via the OP techniques discussed in Sec. \ref{sec:exptechniques}, the number of atoms per spin component $N_S$ being always the same regardless of $N$. The interaction between atoms in different spin states is repulsive, with a 3D scattering length $a \simeq 200 a_0$. 

\begin{figure}[t!]
\begin{center}
\includegraphics[width=0.7\columnwidth]{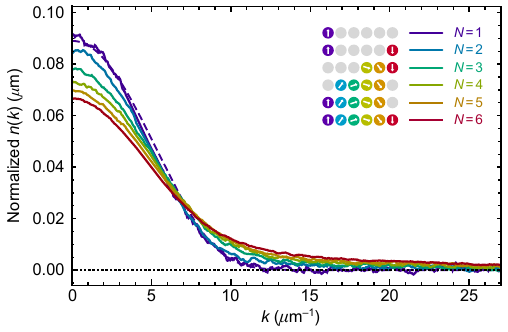}
\end{center}
\caption{Momentum distribution $n(k)$ measured with time-of-flight absorption imaging for different 1D spin mixtures of $^{173}$Yb fermions, with variable number of components $N$ and same number of particles per spin component. The dashed line is the prediction for the noninteracting single-component gas. The experimental curves have been normalized in such a way as to have $\int n(k) dk = 1$. Adapted from Ref. \cite{pagano2014}.} \label{fig:1d-1}
\end{figure}

In Fig. \ref{fig:1d-1} the momentum distribution $n(k)$ of multicomponent SU($N$) gases, as obtained via time-of-flight absorption imaging, is shown. When $N=1$ the system is spin-polarized and it behaves as a noninteracting system, as it can be verified by the very good agreement of the measured $n(k)$ with the prediction for an ideal Fermi gas (dashed line) with no fit parameters. When $N\geq 2$ the momentum distribution is broadened by the repulsion between the particles, with deviations from the ideal gas theory that are more and more important as the number of components is increased. This broadening of the momentum distribution can be understood as a result of the enhanced correlations: from a qualitative point of view, repulsive interactions force atoms to be more localized in position space, in order to reduce the overlap of their wavefunctions. In the limit of infinite repulsion, a phenomenon called ``fermionization" occurs in 1D: the repulsion is so strong that particles cannot occupy the same position in space, this mimicking an effective Pauli repulsion among them. As a consequence, a ``fermionized" spin mixture of $N$ components and $N_s$ atoms per component, regardless of their bosonic or fermionic nature, will exhibit properties of a Fermi gas of $N \times N_s$ particles\footnote{This correspondence does not apply exactly to {\it all} the properties of the system: while collective excitation energies are predicted to be exactly those of an ideal Fermi gas with $N \times N_s$ particles, the momentum distribution function is affected in a more complex way. For more details the reader can refer to specialized reviews \cite{guan2013}.}. Because of the increased localization in coordinate space, the wavefunction will be more delocalized in momentum space. From a quantitative point of view, understanding how this delocalization  occurs is a very hard task. Recent theoretical works \cite{decamp2016} calculated the momentum distribution function expected for SU($N$) Fermi gases, with analytical predictions for the long-$k$ behavior $n(k)\simeq C k^{-4}$. The amplitude $C$ of the long-$k$ tail is called {\it Tan's contact} \cite{tan2008} and for SU($N$) Fermi gases it was measured recently in Ref. \cite{song2020}.

\begin{figure}[t!]
\begin{center}
\includegraphics[width=1\columnwidth]{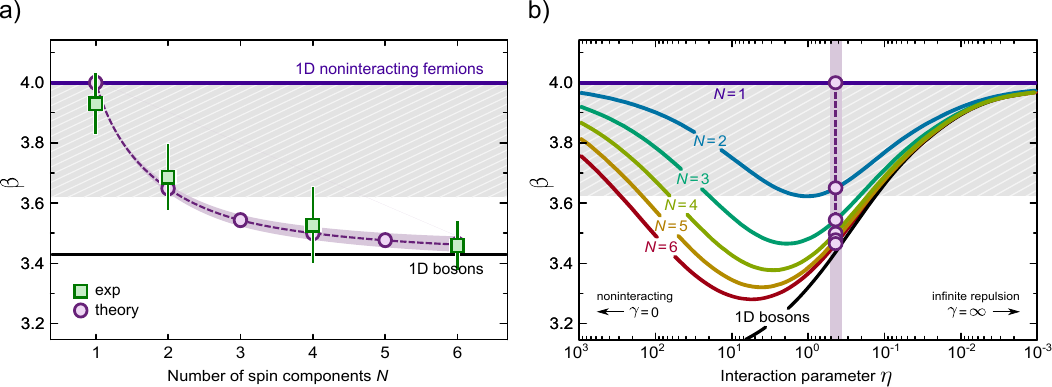}
\end{center}
\caption{Collective excitations of multicomponent 1D spin mixtures of $^{173}$Yb fermions. a) Experimental (squares) and theoretical (circles) values for the squared ratio $\beta$ of the breathing frequency $\omega_B$ to the harmonic trap frequency $\omega$, as a function of the number of spin components $N$; b) theoretical curves showing the large-spin bosonization effect, with the curve for 1D bosons emerging as the limiting curve of the family of 1D fermionic curves for $N \rightarrow \infty$. Adapted from Ref. \cite{pagano2014}.} \label{fig:1d-2}
\end{figure}

Fig. \ref{fig:1d-2} shows the result of a different experiment, where collective excitations of the 1D systems were studied as a function of the number of spin components in the mixture. The lowest-frequency mode for trapped 1D systems is the breathing mode, describing  a periodic oscillation of the system size at frequency $\omega_B$. It can be detected by suddenly quenching the frequency of the harmonic trap $\omega$ and then measuring the time evolution of the cloud size. Fig. \ref{fig:1d-2}a shows the measured breathing frequency, expressed via the parameter $\beta=(\omega_B/\omega)^2$, for different number of spin components $N$. Clearly the experimental data (squares) show a dependence of the breathing mode frequency on $N$: when $N=1$ the experimental value agrees with the theoretical prediction $\beta=4$ for a spin-polarized 1D Fermi gas \cite{astrakharchik2004}, then for $N\geq 2$ we observe a monotonic decrease of the breathing frequency, which approaches the value calculated for 1D bosons for the largest value $N=6$. This effect is a first experimental demonstration of the ``large-spin bosonization" originally predicted in a work by Nobel Prize winner C. N. Yang {et al.} \cite{yang2011}, who demonstrated that a multicomponent 1D fermionic mixture, at zero temperature and in the limit of an infinite number of spin components, exhibits the same Bethe Ansatz solution as a 1D system of bosons, at any interaction strength. This behavior is well illustrated in the theoretical curves calculated by H. Hu and X.-J. Liu, shown in Fig. \ref{fig:1d-2}b \cite{pagano2014}. This is a striking example of the non-intuitive quantum effects that happen in 1D, where quantum correlations induced by interactions become maximally important and the very notion of quantum statistics becomes somewhat ``blurred". Indeed, multicomponent SU($N$) spin mixtures are a very versatile platform that allows for a precise ``tuning" of the quantum distinguishability of the particles and of the role of the Pauli exclusion principle in the emerging physical properties (see also Ref. \cite{song2020} for related experimental studies with three-dimensional spin mixtures).


\section{Experiments with interacting SU($N$) mixtures in optical lattices}
\label{sec:exp-su(n)-lattices}

In this section I will discuss some aspects of the physics of SU($N$) spin mixtures trapped in optical lattices, focusing on the SU($N$) Fermi-Hubbard model and then discussing the effects of an explicit, tunable breaking of the SU($N$) symmetry.

\subsection{SU($N$) Fermi-Hubbard model}

The Fermi-Hubbard model is one of the most studied models in condensed-matter physics: it is the minimal conduction model describing quantum correlations between interacting electrons and finds significant applications in the description of important classes of materials, such as cuprates, that exhibit high-temperature superconductivity. The Fermi-Hubbard model describes a gas of interacting spin-1/2 electrons hopping in a lattice with an extension of the single-band tight-binding Hamiltonian already introduced in Eq. (\ref{eq:tightbinding}):
\begin{equation}
\hat{H} = -t \sum_{\langle i,j \rangle,m}  \hat{c}_{im}^\dagger \hat{c}_{jm}^{ }   + U \sum_{i} \hat{n}_{i \uparrow} \hat{n}_{i \downarrow} \, , \label{eq:fermihubbard}  
\end{equation}
where $m=\{\uparrow,\downarrow\}$ indicates the electron spin state, $\hat{n}_{im}= \hat{c}_{im}^\dagger  \hat{c}_{im}^{ }$ is the number operator counting electrons with spin $m$ in site $i$, and $U$ is the interaction energy associated to the occupation of the same site by two electrons -- one in state $\uparrow$ and one in state $\downarrow$ -- also dubbed as a \textit{doublon}. In the repulsive case $U>0$ and for densities corresponding to the ``half-filling" configuration $\langle n_{i\uparrow} \rangle = \langle n_{i\downarrow} \rangle = 1/2$, this model hosts a quantum phase transition driven by the $U/t$ parameter. When $U \ll t$ the many-body ground state is a metallic state, with delocalized electrons partially filling the Bloch band and nonzero particle number fluctuations (i.e., a finite probability of forming doublons). Instead, when $U \gg t$ the large energetic cost associated to the presence of two electrons in the same site prohibits the formation of doublons and the electrons localize each at an individual lattice site with vanishing particle number fluctuations.

The Fermi-Hubbard model has been the focus of very intense investigations with ultracold atoms in optical lattices \cite{esslinger2010,tarruell2018}. The transition from a metal to a Mott insulator was observed by detecting the formation of doublons \cite{jordens2008}, by measuring the system compressibility \cite{schneider2008} and by direct in-situ imaging \cite{greif2016}. Among the different research directions connected with the quantum simulation of this model, we mention the characterization of the low-temperature magnetic properties of the Mott state. Antiferromagnetic correlations between nearest-neighboring sites were detected \cite{greif2013}, with the eventual observation of long-range antiferromagnetic ordering \cite{mazurenko2017}. Current efforts are aimed at exploring different properties of this model, and in particular at the study of its low-temperature phase at finite hole doping, when a superconducting state is expected to form.

A growing interest, from both the theoretical and experimental point of view, is given to multicomponent Hubbard models, where the internal state of the particles spans a Hilbert space with dimension larger than 2. This is the case of large-spin Hubbard models (when the spin of the particles is larger than 1/2) or multi-orbital Hubbard models (where other quantum numbers are considered in addition to the particle spin). An example of the former class is the SU($N$)-symmetric Fermi-Hubbard model \cite{honerkamp2004}
\begin{equation}
\hat{H} = -t \sum_{\langle i,j \rangle,m} \hat{c}_{im}^\dagger \hat{c}_{jm}^{ }  + \frac{U}{2} \sum_{i,m,m'\neq m} \hat{n}_{i m} \hat{n}_{i m'} \, , \label{eq:sunfermihubbard}  
\end{equation}
where $m=\left\{1,\ldots,N\right\}$. This model, describing spin-$F$ particles with $F=(N-1)/2$ and spin-independent interactions, is characterized by a global SU($N$) symmetry since both the hopping energy $t$ and the interaction energy $U$ do not depend on $m$. Similarly to the SU(2) Fermi-Hubbard model in Eq. (\ref{eq:fermihubbard}), also the SU($N$) Fermi-Hubbard model hosts a quantum phase transition from a metallic to a Mott insulating state when the interaction energy is repulsive. This phase transition occurs at fractional fillings $q/N$, where $q$ is an integer, with the $1/N$ filling corresponding to the case of 1 particle/site.

An intense theoretical effort is devoted to understanding the low-temperature properties of such model, as different kinds of magnetic ordering in the low-temperature Mott phase are possible, as a consequence of the enlarged symmetry. Depending on $N$ and on the lattice geometry, a variety of magnetic phases such as SU($N$) antiferromagnets, SU($N$) resonating-valence-bond states and spin-liquid states have been predicted (see the introduction of Ref. \cite{taie2022} for a partial list of relevant references). From the experimental point of view, the SU($N$) Fermi-Hubbard model was realized in quantum simulators with two-electron atoms, with the characterization of SU($N$) Mott insulators \cite{taie2012,hofrichter2016} and the recent observation of SU($N$) antiferromagnetism in the Mott phase \cite{ozawa2018,taie2022}.

\subsection{Flavour-selective localization in an SU(3)-broken Fermi-Hubbard model}
\label{sec:exp-su(n)-flavourmott}

\begin{figure}[t!]
\begin{center}
\includegraphics[width=1\columnwidth]{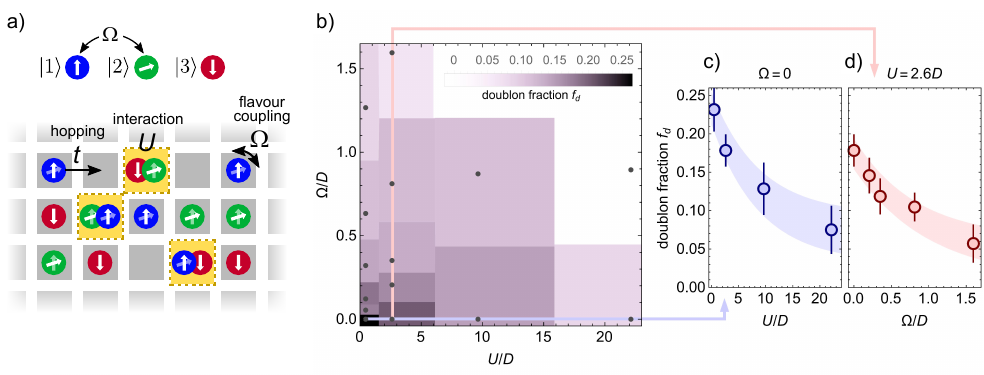}
\end{center}
\caption{a) Sketch of a system of SU(3) interacting fermions in a lattice, where the global symmetry is explicitly broken by a coherent Raman coupling $\Omega$ between two internal states. b) Experimental phase diagram showing the fraction of atoms in doubly occupied sites as a function of atom repulsion $U$ and Raman coupling $\Omega$. c) Subsets of the data for two different cross sections of the plot in b), i.e., for $\Omega=0$ and for $U=2.6D$. Adapted from Ref. \cite{tusi2022}.} \label{fig:mott-1}
\end{figure}

It is quite interesting to study how the properties of the SU($N$) Fermi-Hubbard model in Eq. (\ref{eq:sunfermihubbard}) are modified when the global symmetry of the Hamiltonian is \textit{explicitly} broken, e.g. with the addition of external fields coupling some of the internal states. This is not just an intellectual exercise, rather it finds interesting applications, as modified Hubbard models including internal-state coupling (and/or additional degrees of freedom) have been considered for the description of multi-orbital strongly correlated materials that cannot be described in terms of a plain single-band Hubbard Hamiltonian. These systems are not merely more complicated but rather host new phenomena, challenging the standard paradigm of Mott localization \cite{demedici2009,georges2013}. One of the simplest examples is the following SU(3)-broken Hamiltonian
\begin{equation}
\hat{H} = -t \sum_{\langle i,j \rangle,m} \hat{c}_{im}^\dagger \hat{c}_{jm}^{ }   + \frac{U}{2} \sum_{i,m,m'\neq m} \hat{n}_{i m} \hat{n}_{i m'} + \frac{\Omega}{2} \sum_{i} \left( \hat{c}_{i1}^\dagger \hat{c}_{i2}^{ } + \hat{c}_{i2}^\dagger \hat{c}_{i1}^{ } \right) \, , \label{eq:su3brokenfermihubbard}  
\end{equation}
with $m=\left\{1,2,3\right\}$, where the third term describes a coherent coupling between two specific internal states $m=1$ and $m=2$ (see sketch in Fig. \ref{fig:mott-1}a). This term explicitly breaks the SU(3) symmetry of the Hamiltonian, by an amount which is controlled by the new energy scale $\Omega$.

In recent experiments performed in Florence, following the theoretical proposal of Ref. \cite{delre2018}, we have performed a quantum simulation of the Hamiltonian in Eq. (\ref{eq:su3brokenfermihubbard}) by trapping $^{173}$Yb atoms in a 3D optical lattice \cite{tusi2022}. The atoms are prepared in a subset of three spin states with the optical pumping techniques described in Sec. \ref{sec:exptechniques} and the last term of Eq. (\ref{eq:su3brokenfermihubbard}) is implemented with a two-photon Raman coupling, as described in the same section\footnote{The Raman coupling only connects states $m=1$ and $m=2$, while $m=3$ is left uncoupled, thanks to the control of the light shifts on the Raman transition. For specific details on the experimental protocol see Ref. \cite{mancini2015}.}. Fig. \ref{fig:mott-1}b shows the average number of doublons after an adiabatic state preparation protocol, as a function of $U$ and $\Omega$. The data clearly reveal the cooperative effect of Rabi coupling and repulsive interactions driving the system toward a Mott localized state with a suppressed number of doublons. The same data are plotted with error bars in Fig. \ref{fig:mott-1}c,d along two different line cuts of the diagram in Fig. \ref{fig:mott-1}b. Fig. \ref{fig:mott-1}c shows the effect of an increasing $U$ in the transition toward an SU(3) Mott insulator for $\Omega=0$, while Fig. \ref{fig:mott-1}d shows a similar localization effect induced by increasing $\Omega$ at a fixed interaction strength $U=2.6D$ (where $D= zt$, with $z$ lattice coordination number).

\begin{figure}[t!]
\begin{center}
\includegraphics[width=0.9\columnwidth]{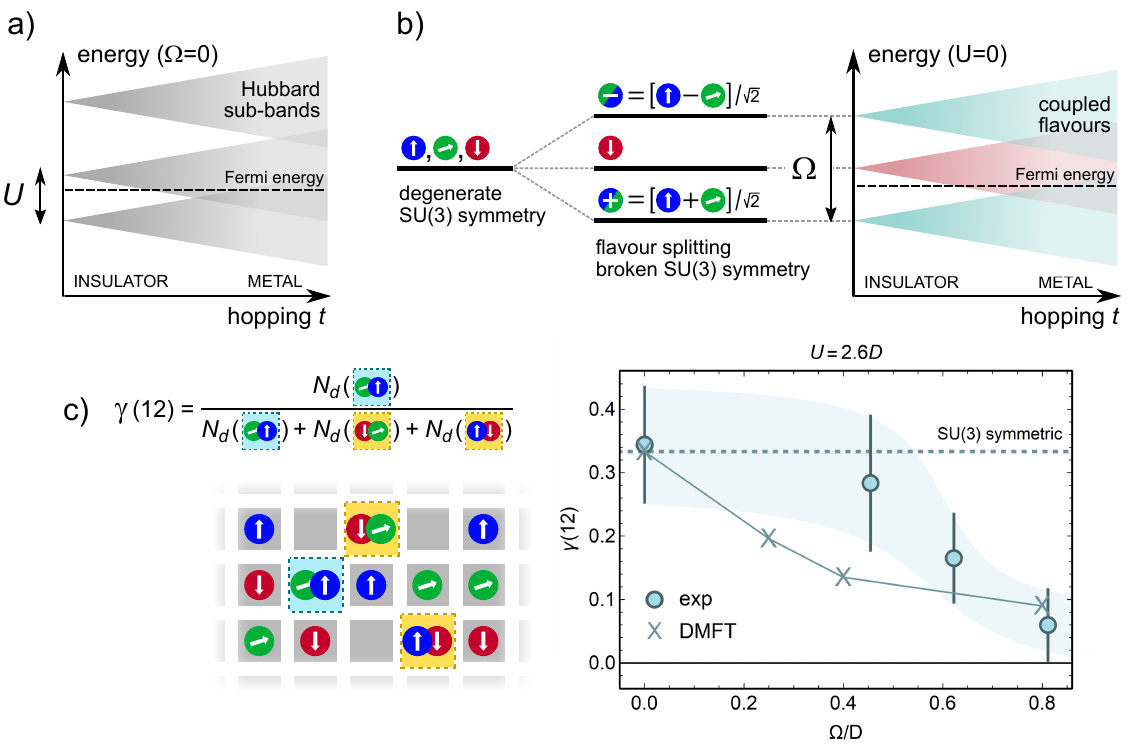}
\end{center}
\caption{a) Sketch of the Mott-metal transition in the symmetric SU(3) Fermi-Hubbard model: when Hubbard sub-bands overlap at large $t$ the metallic behavior is restored. b) The selective Raman coupling lifts the degeneracy between flavours: the competition with the hopping can drive a transition from a metal to an insulator already in the non-interacting case, similarly to the Mott localization scenario. c) Fraction of doublons $\gamma(12)$ formed by the Raman-coupled states (circles are experimental data, crosses are zero-temperature DMFT calculations, the dashed line shows the expected value for a system with SU(3) symmetry). Adapted from Ref. \cite{tusi2022}.} \label{fig:mott-2}
\end{figure}

How to understand, on qualitative grounds, this enhancement of localization by the symmetry-breaking $\Omega$ term? The sketch in Fig. \ref{fig:mott-2}a illustrates the origin of the Mott-metal phase transition for $\Omega=0$: when $t=0$ (atomic limit) the system can be described in terms of discrete energy levels corresponding to different integer number of particles in a site; when $t>0$ these discrete levels become Hubbard bands of width $\sim 2D$  and, when these bands eventually overlap for $D \simeq U$, particle number fluctuations set in and the metallic behavior is recovered. A similar picture, shown in Fig. \ref{fig:mott-2}b for the simple $U=0$ case, can be used to understand the effect of the symmetry-breaking coupling $\Omega$ on the localization  of the system. When $\Omega \neq 0$ the degeneracy among the three flavours is lifted, as $|1\rangle$ and $|2\rangle$ form two dressed states $|\pm\rangle = (|1\rangle \pm |2\rangle)/\sqrt{2}$, which are energy-shifted from $|3\rangle$: again, when hopping is introduced, these discrete energy levels become energy bands, similarly to what is sketched in Fig. \ref{fig:mott-2}a, and a transition from an insulating state (with suppressed particle fluctuations) at $D=0$ to a metallic state when $D \simeq \Omega$ occurs\footnote{In this simple non-interacting picture, the localized state for $D \ll \Omega$ is actually a band insulator (all the sites are occupied by atoms in the same flavours)}. 

The experimental findings are well supported by a numerical solution of the model in Eq. (\ref{eq:su3brokenfermihubbard}) \cite{tusi2022}. Theory also predicts the onset of \textit{flavour-dependent correlations}, which may eventually lead to the formation of flavour-selective Mott insulating states, where only atoms in specific flavours (or combinations of them) are localized, while the others have a metallic nature. This flavour-selective behaviour can be detected experimentally by resolving the spin character of the doublons, i.e., by counting how many atoms form doublons in each of the three pairs $|12\rangle$, $|23\rangle$ and $|31\rangle$. Fig. \ref{fig:mott-2}c shows the quantity $\gamma(12)=N_d(12)/N_d$, where $N_d(mn)$ is the number of atoms forming doublons in the $|mn\rangle$ channel, as a function of $\Omega$ and fixed $U=2.6D$. The measured value at $\Omega=0$ agrees with the expectation $\gamma(12)=1/3$ for the SU(3)-symmetric case. As $\Omega$ is increased and the SU(3) symmetry is broken, $\gamma(12)$ diminishes, eventually approaching zero for $\Omega \approx D$. The doublons acquire a strong flavour-selective behaviour, with state $|3\rangle$ showing a higher likelihood to form doublons, which points out at a larger mobility of the particles in this state, compared to $|1\rangle$ and $|2\rangle$. This suppression of $|12\rangle$ doublons is triggered by the polarization effect in the internal-state basis, which can be understood, at a qualitative level, already from the non-interacting picture of Fig. \ref{fig:mott-2}c. While $|23\rangle$ and $|31\rangle$ doublons can be formed by two fermions in the lowest single-particle states $|+\rangle$ and $|3\rangle$, $|12\rangle$ doublons can be formed only if the two fermions occupy states $|+\rangle$ and $|-\rangle$, therefore with an additional energy cost of $\Omega/2$. This mechanism, here described in a simplified non-interacting scenario, does provide the trigger for the onset of flavour-selective correlations in the Mott localized phase. DMFT calculations realized by the group of M. Capone at SISSA confirm the trend of the experimental data, as shown in Fig. \ref{fig:mott-2}c (deviations can be attributed to the effect of finite-temperature, not included in the theory).

The two observations reported above -- the enhancement of Mott localization by flavour coupling and the onset of flavour-selective correlations -- are the two key signatures of selective Mott physics \cite{delre2018}. This concept is a generalization of the ``orbital-selective" Mott scenario, which has become a central paradigm for the description of high-Tc iron-based superconductors, as it can describe the anomalies of the metallic state \cite{demedici2014} and the orbital character of superconductivity \cite{sprau2017} in those systems. Extensions of our investigations could shed light on the behavior of these novel materials and on the rich many-body physics exhibited by the orbital Hubbard models employed for their description.


\section{Multicomponent systems and synthetic dimensions}
\label{sec:exp-synthetic}

\begin{figure}[t!]
\begin{center}
\includegraphics[width=0.76\columnwidth]{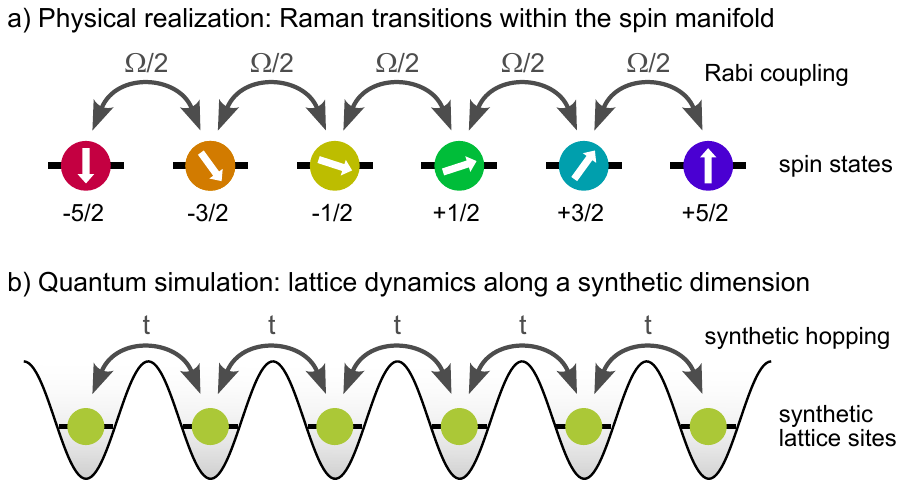}
\end{center}
\caption{a) Sketch of the laboratory implementation of Raman couplings between nearest-neighboring spin projection states. b) Quantum-simulation interpretation of the spin dynamics in terms of quantum hopping between sites along a fictitious synthetic dimension.} \label{fig:synth}
\end{figure}

The experimental realization of spin mixtures with tunable coherent coupling allowed for the elaboration of an intriguing concept for the design of novel quantum simulation schemes, that of {\it synthetic dimensions}. To explain what a synthetic dimension is, let's reconsider what happens in the nuclear-spin manifold under the influence of the coherent two-photon Raman coupling already described in Secs. \ref{sec:exptechniques} and \ref{sec:exp-su(n)-flavourmott}. Fig.  \ref{fig:synth}a shows a sketch of the effective Raman coupling induced by two laser fields with different polarization, inducing $\Delta m = \pm 1$ transitions. The Hamiltonian can be written as:
\begin{equation}
\hat{H} =   \frac{\Omega}{2} \sum_{\langle m,m' \rangle} \hat{c}_m^\dagger \hat{c}_{m'}^{ }  \, , \label{eq:synthdim-tightbinding}  
\end{equation}
where $\langle m,m' \rangle$ denotes ``nearest-neighbor" spin states with projection numbers $m$ and $m'=m \pm  1$, and $\Omega$ is the Rabi frequency describing the strength of the coupling\footnote{The Rabi frequency 
$\Omega$ depends on the laser intensities and detuning from the excited electronic state, according to Eq. (\ref{eq:rabiraman}). We note that, because of the Clebsch-Gordan coefficients entering the description of laser-matter interaction, $\Omega$ generally depends on $m$ and $m'$ (for more details see e.g. Ref. \cite{steck2022}). For the sake of simplicity we will ignore this dependence in this introductory paragraph.}. 

The similarity of this Hamiltonian with the tight-binding lattice model in Eq. (\ref{eq:tightbinding}) is apparent. The two equations can be mapped one into the other by just replacing the spin quantum numbers $m,m'$ and the Raman amplitude $\Omega/2$ in Eq. (\ref{eq:synthdim-tightbinding}) with the lattice site indexes $i,j$ and the tunnelling strength $t$, respectively, in Eq. (\ref{eq:tightbinding}). This analogy led the authors of Ref. \cite{boada2012} to develop the concept of a synthetic dimension, by interpreting the dynamics in the {\it internal} Hilbert space spanned by the atomic spin as if it described the quantum motion of a particle along a {\it fictitious} dimension with an embedded lattice structure (as sketched in Fig. \ref{fig:synth}b).

The idea of synthetic dimensions is completely general. It just requires the experimental access to a quantum degree of freedom with coherent coupling between its quantum states. The degree of freedom used for the simulation of the lattice dynamics can be the spin of an atom, its electronic state, the atomic momentum, the quantum levels of the atom confined in a trap, the rotational levels of a molecule, etc... A comprehensive review of proposals and experiments can be found in Ref. \cite{ozawa2019}.

\subsection{Synthetic dimensions and artificial magnetic fields}

\begin{figure}[t!]
\begin{center}
\includegraphics[width=\columnwidth]{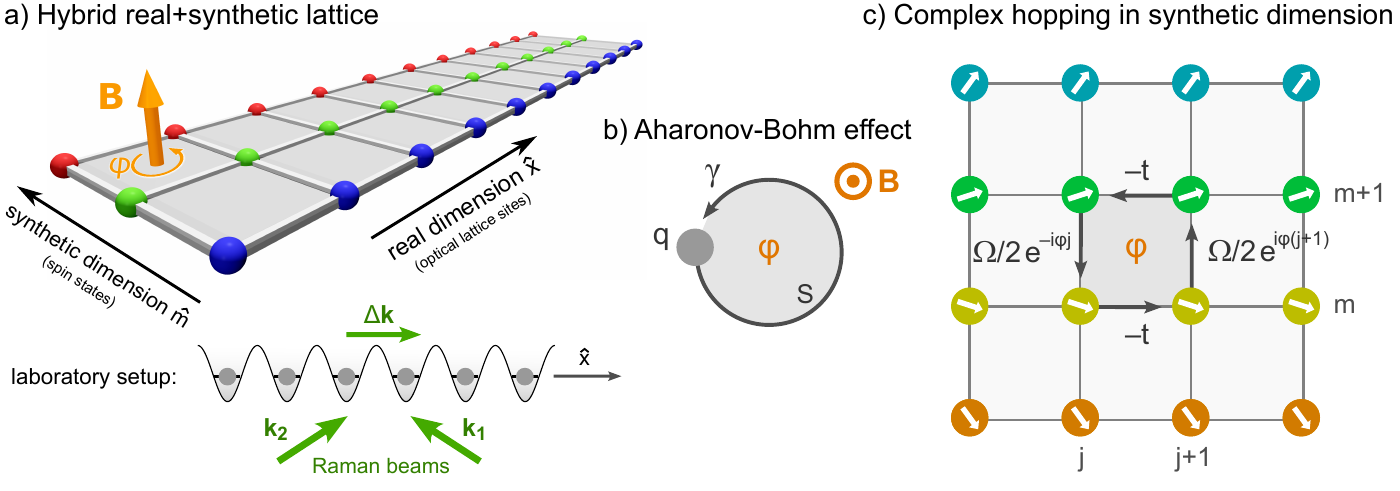}
\end{center}
\caption{a) Sketch of the laboratory realization of a hybrid real + synthetic lattice.  b) In the Aharonov-Bohm effect the wavefunction acquires a nonzero phase when a charged particle moves around a closed loop in the presence of an enclosed magnetic field. c) The position-dependent phase of the Raman coupling produces the equivalent of an Aharonov-Bohm phase when a neutral atom hops around a unit cell of the hybrid lattice. } \label{fig:synth2}
\end{figure}

The concept of synthetic dimensions found a first practical application in the investigation of topics of topological quantum physics. In Ref. \cite{celi2014} it was realized that the combination of a standard optical lattice with a synthetic-dimensional lattice could allow for the simplest experimental method to engineer synthetic magnetic fields for particles with an effective charge  (see Refs. \cite{dalibard2011,goldman2014} for an introduction to the field and comprehensive reviews of different experimental techniques). The main idea is based on the synthesis of an effective Aharonov-Bohm phase for an atom encircling a unit cell of a hybrid two-dimensional lattice, where one dimension is real and one is synthetic, as represented in Fig. \ref{fig:synth2}a. 

In quantum mechanics the Aharonov-Bohm phase is the result of the action of a magnetic field onto a charged particle. We recall that in free space it can be derived from the Hamiltonian for a particle of charge $q$ in a magnetic field $\mathbf{B}=\nabla \times \mathbf{A}$ described by a magnetic vector potential $\mathbf{A}(\mathbf{r})$
\begin{equation}
\hat{H}=\frac{1}{2M}\left( \mathbf{p} - q \mathbf{A}(\mathbf{r})\right)^2 + V(\mathbf{r}) \; ,
\label{eq:h-magneticfield}
\end{equation}
where $\mathbf{p}$ is the canonical momentum and $V(\mathbf{r})$ is a scalar potential. It is easy to verify that the general form of the wavefunction $\psi(\mathbf{r},t)$ that solves the time-dependent Schr{\"o}dinger equation with the Hamiltonian in Eq. (\ref{eq:h-magneticfield}) is given by
\begin{equation}
\psi(\mathbf{r},t)=\psi_0(\mathbf{r},t) e^{ \frac{\mathrm{i}q}{\hbar} \int_0^\mathbf{r} \mathbf{A}(\mathbf{r'}) \cdot d\mathbf{r'} } \; ,
\end{equation}
where $\psi_0(\mathbf{r},t)$ is the solution for $\mathbf{A}=0$: the effect of the magnetic field is a geometric phase shift on the wavefunction. Following the semiclassical argument used in standard quantum-mechanics books to derive the Aharonov-Bohm interference effect, it is easy to show that, for a particle moving in a closed loop $\gamma$ in the presence of a magnetic field (let's suppose it uniform for the sake of simplicity), when the particle comes back to the original position (Fig. \ref{fig:synth2}b) its wavefunction gets a phase shift
\begin{equation}
\varphi = \frac{q}{\hbar} \oint_\gamma \mathbf{A}(\mathbf{r'}) \cdot d\mathbf{r'} = \frac{q}{\hbar} \iint_S \mathbf{B} \cdot d\mathbf{S} = 2\pi \frac{\Phi_S(\mathbf{B})}{\Phi_0} \; ,
\label{eq:aharonovbohmphase}
\end{equation}
where $\Phi_S(\mathbf{B})$ is the flux of the magnetic field through the surface $S$ enclosed by $\gamma$ and $\Phi_0=h/q$ is the quantum of flux\footnote{In the second equivalence of Eq. (\ref{eq:aharonovbohmphase})  the Stokes' theorem has been used.}.

Peierls \cite{peierls1933} showed that, for a quantum particle moving in a lattice in the presence of a magnetic vector potential $\mathbf{A}(\mathbf{r})$, the corresponding tight-binding Hamiltonian, usually called the \textit{Harper-Hofstadter model} \cite{harper1955,hofstadter1976}, can be written as a modification of the Hamiltonian in Eq. (\ref{eq:tightbinding}),
\begin{equation}
\hat{H} = -t \sum_{\langle i,j \rangle} e^{\mathrm{i} \theta_{ji}} \hat{c}_i^\dagger \hat{c}_j^{ }  \, ,
\label{eq:tightbindingpeierls}  
\end{equation}
where the hopping amplitudes are now complex values with phases $\theta_{ji} = (q/\hbar) \int_\mathbf{r_j}^\mathbf{r_i} \mathbf{A}(\mathbf{r'}) \cdot d\mathbf{r'}$, usually called {\it Peierls phases}. It is immediate to note that the sum of the Peierls phases around a unit cell of the lattice corresponds to the dimensionless flux of magnetic field piercing the cell $\varphi$, according to Eq. (\ref{eq:aharonovbohmphase}).

Peierls phases can be conveniently synthesized taking advantage of the position-dependent phase of the Raman coupling realizing the hopping in the synthetic dimension. Indeed, when the Raman beams propagate along different real-space directions $\mathbf{k_1}$ and $\mathbf{k_2}$ (see Fig. \ref{fig:synth2}a), the two-photon Rabi frequency in Eq. (\ref{eq:rabiraman}) acquires a space-dependent phase factor. Recalling that the single-photon Rabi frequency $\Omega_i$ is proportional to the electric field $E_i=E_{0i} \exp(i  \mathbf{k_i}\cdot\mathbf{r})$, we have 
\begin{equation}
    \Omega = \frac{\Omega_1^* \Omega_2} {2 \delta} = \frac{\left(
    E^*_{01} e^{-\mathrm{i}  \mathbf{k_1}\cdot\mathbf{r}}
    \right)\left(
    E_{02} e^{\mathrm{i}  \mathbf{k_2}\cdot\mathbf{r}}
    \right)} {2 \delta} =
    |\Omega| e^{\mathrm{i}  \left( \mathbf{k_2}-\mathbf{k_1}\right) \cdot\mathbf{r}}
    \; ,
    \label{eq:rabiramanphase}
\end{equation}
so the hopping along the synthetic dimension comes with a phase factor that depends on the real-space position. Assuming the Raman beams to be oriented with their wavevector difference $\Delta \mathbf{k} = \mathbf{k}_2-\mathbf{k}_1$ lying along the real-lattice direction, we can write the position-dependent Rabi frequency as 
\begin{equation}
    \Omega =  |\Omega| e^{\mathrm{i}  \varphi j}  ,
    \label{eq:rabiramanphase2}
\end{equation}
where $j$ is the site index and $\varphi = d |\Delta \mathbf{k}|$, with $d$ the real-lattice spacing.

In Fig. \ref{fig:synth2}c the hopping matrix elements around a unit cell of the hybrid real+synthetic lattice are represented. It is clear that the sum of the Peierls phases is just $\varphi$, independently of the specific cell under consideration, as if the lattice were pierced by a uniform dimensionless magnetic flux $\varphi$ per unit cell.

\subsection{Experimental observation of chiral edge currents in synthetic ladders}
\label{sec:exp-synthetic-currents}

Synthetic dimensions were first realized in experiments performed in 2015 in Florence and at NIST/JQI \cite{mancini2015, stuhl2015}, where the idea for the generation of synthetic artificial magnetic fields described in the previous section \cite{celi2014} was realized. 

\begin{figure}[t!]
\begin{center}
\includegraphics[width=\columnwidth]{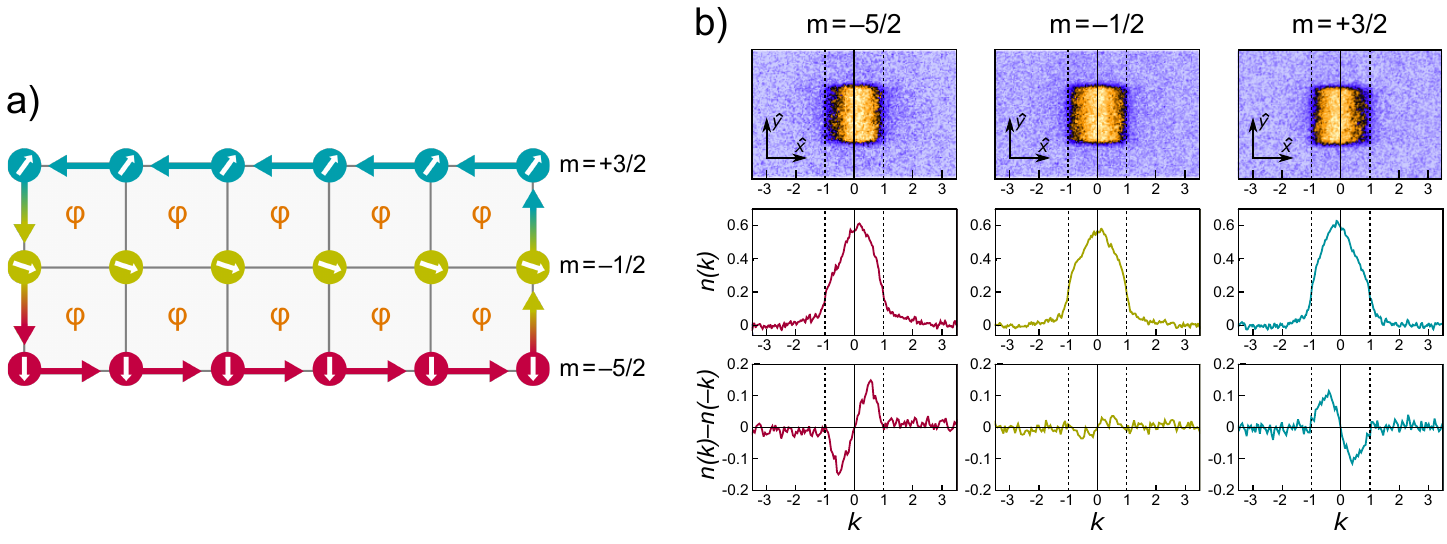}
\end{center}
\caption{a) Sketch of chiral edge currents for a hybrid real + synthetic ladder pierced by a synthetic flux $\varphi$ per unit cell: atoms in the highest leg ($m=+3/2$) and in the lowest leg ($m=-5/2$) move in opposite direction.  b) Top: experimental images obtained with spin-selective time-of-flight detection. Center: lattice momentum distribution $n(k)$ along the real-space direction $\hat{x}$  for each leg of the synthetic ladder. Bottom: ``asymmetry" function $n(k)-n(-k)$ evidencing the asymmetric distributions for the highest and lowest leg. The synthetic magnetic flux is $\varphi\simeq 0.37 \pi$. 
 Adapted from Ref. \cite{mancini2015}.} \label{fig:synth3}
\end{figure}

The main result reported in these experiments is the observation of chiral edge currents. Edge currents are a hallmark of topological states of matter, as it happens e.g. in topological insulators, which are insulating in the bulk and can sustain currents only along the edges. In a quantum Hall state these edge modes are chiral, i.e., the direction of the current (clockwise or counterclockwise) is set by the direction of the magnetic field, which breaks time-reversal invariance \cite{halperin1982}. In the synthetic dimension realization, visualizing chiral currents is particularly simple because of two reasons: 1) the system naturally realizes a \textit{ladder} geometry, i.e., the synthetic dimension just spans a few sites (three for the experiments of Refs. \cite{mancini2015,stuhl2015}), which makes edge effects dominate the system properties; 2) performing spin-selective imaging with the techniques described in Sec. \ref{sec:exptechniques} allows for the ``in-situ" detection of properties along the synthetic dimension, with ``single-rung" resolution. Hence, if a net chiral edge current in the system exists (as sketched in Fig. \ref{fig:synth3}a), it can be observed by probing asymmetries in the spin-selective momentum distribution.

Fig. \ref{fig:synth3}b shows the lattice momentum distribution $n(k)$ of three-leg ladders realized with $^{173}$Yb fermions in the Florence experiment \cite{mancini2015}. Here the Raman coupling is switched on adiabatically in order to prepare the system in an equilibrium, low-energy state. Each column of Fig. \ref{fig:synth3}b refers to a specific spin state (i.e., an individual leg of the ladder), while the rows show, from top to bottom, the time-of-flight images, the reconstructed $n(k)$ and the function $n(k)-n(-k)$, which highlights the asymmetry of the momentum distribution. It is evident that the highest leg and lowest leg of the ladder are occupied by atoms with a nonzero mean momentum, positive for the lowest leg ($m=-5/2$) and negative for the highest leg ($m=+3/2$). The central leg ($m=-1/2$), instead, shows a highly symmetric distribution, corresponding to almost null mean momentum. This is 
a direct observation of chiral edge currents in a fermionic system and a demonstration of the versatility of the synthetic-dimension approach.

\begin{figure}[t!]
\begin{center}
\includegraphics[width=\columnwidth]{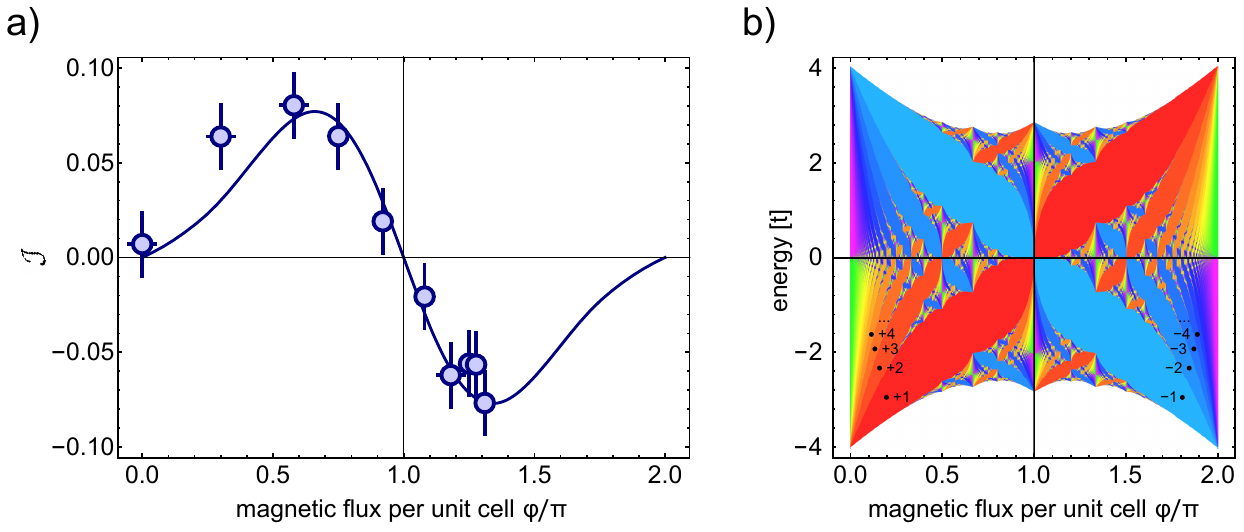}
\end{center}
\caption{a) Measurement of the chiral current imbalance $\mathcal{J}$ as a function of the magnetic field flux for a fermionic two-leg ladder. The line is the theoretical prediction from a noninteracting model. Adapted from Ref. \cite{livi2016}. b) Chern number for the Harper-Hofstadter Hamiltonian for a two-dimensional system with $\Omega=2t$ and periodic boundary conditions: warm colors indicate positive Chern numbers, cold colors indicate negative Chern numbers. The small numbers in the plot are the Chern numbers for the largest energy gaps in the Hofstadter spectrum.} \label{fig:synth4}
\end{figure}

The synthetic magnetic flux can be tuned by changing the direction of the Raman laser beams. Fig. \ref{fig:synth4} shows an experimental measurement of the chiral current for a synthetic two-leg ladder\footnote{Actually, in this experiment the synthetic dimension was realized by exploiting two electronic states, rather than two nuclear-spin states. This is possible in two-electron fermions due to the existence of a metastable state $^3P_0$, which can be excited from the ground state $^1S_0$ with a single-photon ``clock" transition. For more information see Ref. \cite{livi2016} and Sec. \ref{sec:furtherdirections} of these notes.} as a function of the magnetic flux $\varphi$ \cite{livi2016}. The chiral current is here quantified in an effective way by the integrated imbalance of the momentum distribution in a single leg $\mathcal{J}=\int_0^1 (n(k)-n(-k))dk $ \footnote{Here and in Fig. \ref{fig:synth3}b the real-lattice quasimomentum $k$ is defined in units of $\pi/d$, so the integral in the definition of $\mathcal{J}$ spans the positive-$k$ half of the real-lattice first Brillouin zone.} (see bottom plots of Fig. \ref{fig:synth3}b). In the plot of Fig. \ref{fig:synth4}a $\mathcal{J}$ clearly shows a non-monotonic behavior, reaching a maximum, then vanishing at $\varphi=\pi$, then changing sign. Clearly, this behavior cannot be understood in terms of classical physics\footnote{Let's pretend we don't know quantum physics and try to find a classical explanation. We could understand the emergence of a chiral edge current as the macroscopic remnant of the many cyclotron orbits described microscopically by each particle (endowed with an effective charge) under the influence of the synthetic magnetic field $\mathbf{B}$. The direction of the cyclotron motion (clockwise or counterclockwise) is fixed by the direction of $\mathbf{B}$ via the Lorentz force $\mathbf{F}=q\mathbf{v} \times \mathbf{B}$. Therefore, classically speaking, it would make no sense for the edge current to invert direction at large magnetic fields: it would imply a failure of the right-hand rule for the Lorentz force, which would instead become a left-hand rule at large fields!}. The explanation of this effect has to be found in the discreteness of space, i.e., we are considering a lattice model where there is a natural length scale (the lattice spacing) and a minimal unit of area (the lattice unit cell) that is used to quantify the flux. Since the effect of the magnetic flux is that of imprinting a geometric phase $\varphi$ on the wavefunction, it is clear that the physics should not change when this phase is increased by $2\pi$ (i.e., when the magnetic flux per unit cell increases by one quantum of flux):
\begin{equation}
\mathcal{J}(\varphi+2\pi)=\mathcal{J}(\varphi) \, .
\end{equation}
The chiral current should also change sign under reversal of the magnetic field flux,
\begin{equation}
\mathcal{J}(-\varphi)=-\mathcal{J}(\varphi) \, ,
\end{equation}
and, combining these two equations, it immediately follows that the chiral current should vanish at $\varphi = \pi$ and then change sign for $\varphi > \pi$.

This effect is also connected with the expected behavior for a truly two-dimensional system described by the Harper-Hofstadter model of Eq. (\ref{eq:tightbindingpeierls}), for which the concepts of topological quantum physics can be conveniently applied. Fig. \ref{fig:synth4}b shows the single-particle spectrum of the model as a function of the magnetic field, the so-called \textit{Hofstadter butterfly}, where the energy gaps between different magnetic sub-bands are colored according to the \textit{Chern number} of the lower-lying bands\footnote{We do not enter here the definition and properties of this quantity. See e.g. Ref. \cite{goldman2016} for an introduction to it in the context of quantum simulation of topological physics.}. The Chern number measures the number and direction of chiral edge modes, as stated by the bulk-boundary correspondence principle \cite{hatsugai1993}. The inversion of chirality observed in Fig. \ref{fig:synth4}a at $\varphi=\pi$ is thus reflected in the change of sign of the Chern number for the full two-dimensional model described by Fig. \ref{fig:synth4}b.

Finally, we note that this inversion of chirality, achieved when the magnetic flux per lattice unit cell  becomes of the same order as the quantum of flux, cannot be achieved in ordinary matter. As a matter of fact, considering a square lattice with lattice spacing $d=5$ \r{A} (representative of a vast class of real solids), the magnetic field that would be needed in order to have the inversion of chirality at $\varphi = \pi$ is $B \simeq 8300$ T, which exceeds the largest magnetic fields currently achievable on Earth by more than two orders of magnitude! This is also why in the standard description of the quantum Hall effect the underlying lattice structure of the solid state is usually not considered.

\subsection{Experimental study of the Hall response in interacting fermions}
\label{sec:exp-synthetic-hall}

The Hall effect \cite{hall1879} is one of the most important effects in solid-state physics. It has a wide range of applications, from the characterization of carrier properties in materials to the development of accurate magnetic sensing devices \cite{popovic2003}. At large magnetic fields, the discovery of the quantum Hall effect \cite{yoshioka2002}, in its integer and fractional versions, was awarded with Nobel Prizes and triggered the development of topological quantum physics 
\cite{haldane2017}. While the Hall effect  is very well understood in the case of noninteracting electrons, for strongly correlated materials strong deviations are observed and no simple theory can describe the experiments, even in the classical Hall regime. Thus, in a quantum-simulation approach, it is highly desirable to engineer synthetic quantum systems featuring a Hall response and to characterize it as a function of the interactions among the particles. 

\begin{figure}[t!]
\begin{center}
\includegraphics[width=\columnwidth]{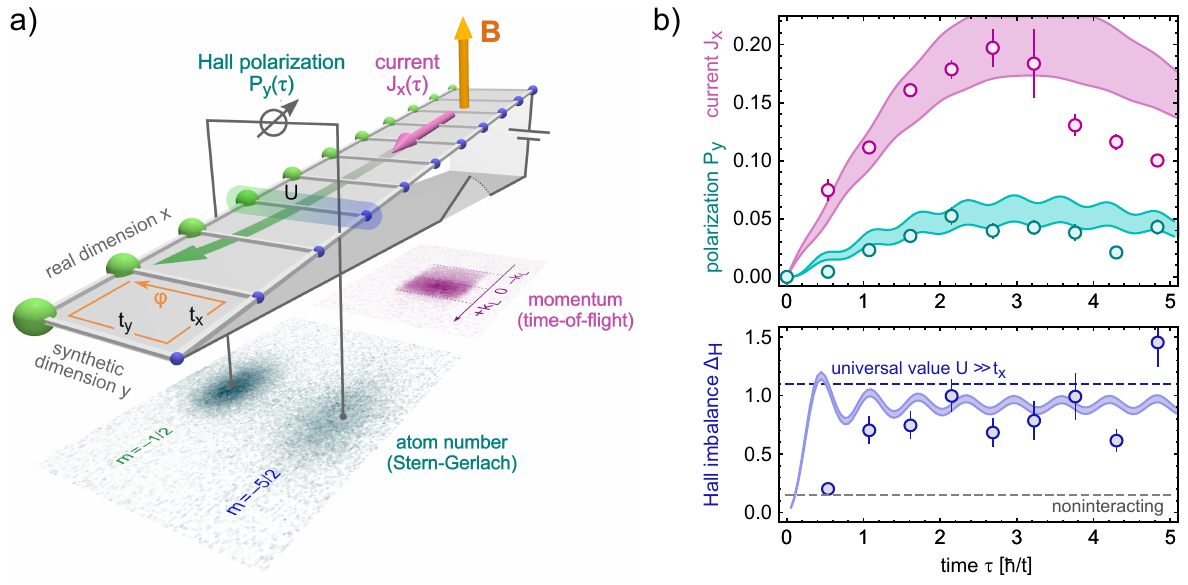}
\end{center}
\caption{Measurement of the Hall response in synthetic fermionic ladders. a) Sketch of the experimental configuration: a potential gradient along real direction $\hat{x}$ induces a current $J_x$ (measured with time-of-flight detection), which is deflected by the Hall drift resulting in a polarization $P_y$ (measured with optical Stern-Gerlach detection). b) Time evolution of $J_x$, $P_y$ and of the Hall imbalance $\Delta_H$ defined in Eq. (\ref{eq:hallimbalance}): the circles are the experimental points and the solid bands are the predictions of an effective model taking into account interactions and finite temperature. The upper horizontal dashed line in the lower panel is the analytic prediction for the universal steady-state value at large interaction strength \cite{greschner2019}, while the lower dashed line is the prediction of a noninteracting model. Adapted from Ref. \cite{zhou2022}.} \label{fig:hall1}
\end{figure}

In Florence, we have performed very recent experiments where we have used the synthetic-dimension approach described in the previous sections to engineer artificial systems where the onset of a Hall response can be  directly observed and characterized \cite{zhou2022}. Fig. \ref{fig:hall1}a shows a sketch of the experimental configuration, i.e., a two-leg ladder with a synthetic magnetic flux similarly to the configuration discussed in the previous section, but with an added potential gradient along the real direction -- the equivalent of a synthetic uniform electric field along $\hat{x}$ -- which induces a longitudinal current $J_x$. Because of the magnetic field, a transverse Hall drift is developed\footnote{In the classical description of the Hall effect, the magnetic field produces a deflection of the carriers according to the Lorentz force $\mathbf{F}=q\mathbf{v}\times\mathbf{B}$, which causes an accumulation of charges on one side of the material. At equilibrium, this charge imbalance produces a transverse electric field that cancels the effect of the Lorentz force on the carriers, thus stopping a further accumulation of charges at the edges.}, which causes an imbalance of population $P_y = N_\uparrow - N_\downarrow$, defined as the population difference in the two legs ($m=-5/2$ and $m=-1/2$ in the experimental realization of Ref. \cite{zhou2022}, here denoted as $\uparrow$ and $\downarrow$). After a sudden activation of the longitudinal potential gradient, these two quantities start evolving  in time, but their ratio
\begin{equation}
\Delta_H = \frac{P_y}{J_x} \, ,
\label{eq:hallimbalance}
\end{equation}
dubbed the \textit{Hall imbalance}, is well defined, as originally noted in the theoretical proposal of Ref. \cite{greschner2019}, and it rapidly reaches a steady-state value, as shown in Fig. \ref{fig:hall1}b \footnote{The quantity $\Delta_H$ is a proxy of the Hall resistivity $E_y/J_x$ that characterizes the Hall effect in real materials and is well suited to describe the Hall response in a relaxation-free setting like that of an atomic quantum simulator.}. 

The two horizontal dashed lines mark the expected Hall response in the noninteracting case and for infinitely strong interactions, for which a universal regime -- independent of atomic density and strength of interactions -- was predicted to exist \cite{greschner2019}. The experimental data, corresponding to an interaction strength $U/t_x=3.28$, clearly deviate from the predictions of the noninteracting theory and approach the universal value $\Delta_H=(2 t_x/t_y)\tan(\phi/2)$ (upper dashed line in Fig. \ref{fig:hall1}b), which only depends on the tunnelling strengths $t_x,t_y$ along the two directions and on the magnetic flux $\phi$ \footnote{The interaction energy $U$ is defined as in the Hubbard model in Eq. (\ref{eq:fermihubbard}). Please note, however, that interactions are local only along the real direction: along the synthetic direction they are {\it non-local}, i.e., they couple all the atoms along the same rung (physically occupying the same real-space position) in an effective infinite-range interaction.}. 

These results, reported in Ref. \cite{zhou2022}, open new perspectives for the quantum simulation of strongly interacting topological systems. Indeed, this is one of the first experiments where a strong effect of atom-atom interactions is observed in a truly many-body system subjected to synthetic gauge fields, in agreement with theoretical predictions. One of the most direct and interesting extensions is the investigation of quantized regimes of the Hall response in systems with larger synthetic-dimension size (i.e., increasing the number of spin components, as recently realized in Ref. \cite{chalopin2020}) and of novel quantum phases recently predicted for synthetic ladders \cite{barbarino2015,taddia2017,calvanese2017}, also in connection with the fractional quantum Hall effect.


\section{Further directions: quantum mixtures of different electronic states}
\label{sec:furtherdirections}

\begin{figure}[t!]
\begin{center}
\includegraphics[width=0.7\columnwidth]{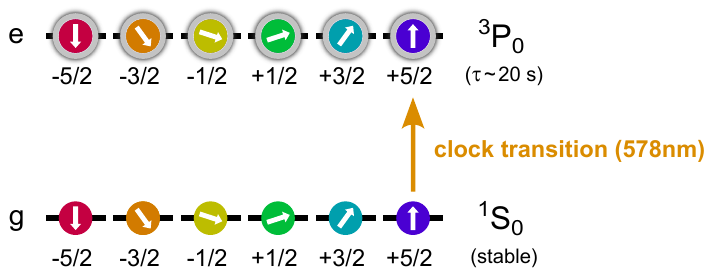}
\end{center}
\caption{In addition to the nuclear-spin states, two-electron fermions such as $^{173}$Yb feature a long-lived electronic state $^3P_0$, which can be accessed from the electronic ground state $^1S_0$ via excitation 
 on an ultranarrow optical clock transition.} \label{fig:electronic}
\end{figure}

During my lectures I focused on selected examples of quantum simulations that were made possible by the optical manipulation of quantum spin mixtures of $^{173}$Yb fermions. In addition to the nuclear-spin degree of freedom, these two-electron atoms are characterized by an \textit{electronic} degree of freedom with metastable states. Among these, we mention the lowest-lying electronic triplet state $|e \rangle$ $=$ $^3P_0$, with a radiative lifetime of $\approx 20$ s, which can be excited starting from the electronic ground state $|g \rangle$ $=$ $^1S_0$ with an ultranarrow optical transition\footnote{The $^1S_0$ $\rightarrow$ $^3P_0$ transition is doubly forbidden, since it apparently violates two selection rules: the rule $\Delta S=0$ that would prohibit transitions between singlet and triplet states, and the rule $J=0 \nrightarrow J'=0$ that would forbid transitions between two states with zero angular momentum. Actually, these transitions can be driven because state $^3P_0$ is weakly admixed to the first excited singlet state $^1P_1$ by spin-orbit coupling and hyperfine interaction. The matrix element is, however, strongly suppressed, which explains the long lifetime of the $^3P_0$ state.} (see Fig. \ref{fig:electronic}). This transition is often called {\it clock} transition, as it is widely used by the metrological community for the realization of optical atomic clocks \cite{ludlow2015}. It can be driven with ultrastable lasers profiting of the advances in laser stabilization technology of the last two decades.

The optical manipulation of the electronic degree of freedom allows for the realization of richer quantum mixtures, with new physical effects and advanced manipulation schemes. For instance, working with atoms in different electronic states allows for the convenient realization of state-dependent optical dipole potentials with far-detuned light and minimal heating effects. It is possible to choose among a full spectrum of possibilities, by just choosing the proper wavelength of the trapping lasers: it is possible to trap both $|g\rangle$ and $|e\rangle$ with the same trapping strength (at the so-called \textit{magic} wavelengths, introduced in the context of optical lattice clocks \cite{derevianko2011}), to trap only one of two, or to trap one and to anti-trap the other.

\subsection{Interorbital spin-exchange dynamics}

\begin{figure}[t!]
\begin{center}
\includegraphics[width=\columnwidth]{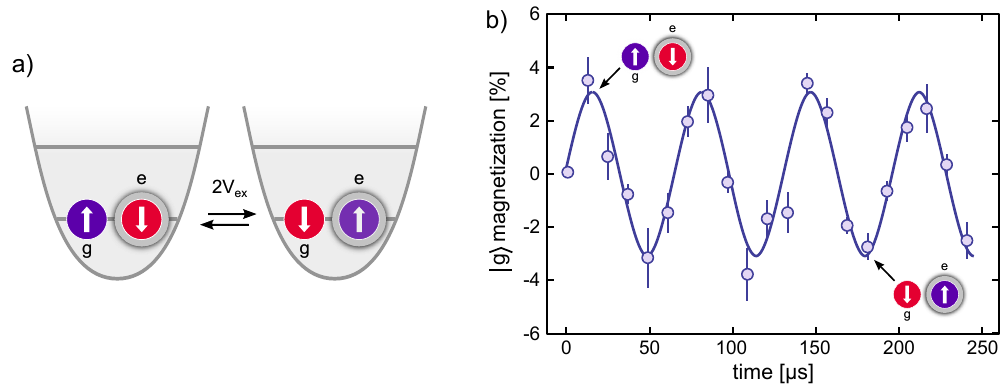}
\end{center}
\caption{Orbital spin-exchange dynamics in $^{173}$Yb fermions. a) When two $^{173}$Yb atoms are trapped with different electronic {\it and} nuclear-spin states, a strong local spin-exchange interaction is observed. b) Measured oscillation of the magnetization of the atoms in the electronic ground state $g$, driven by the orbital spin-exchange interaction $V_{ex}$ in each site of a 3D optical lattice, as described by Eq. (\ref{eq:spinexchange}). Adapted from Ref. \cite{cappellini2014}.} \label{fig:spinexchange}
\end{figure}

The main difference of electronic-state mixtures with respect to nuclear-spin mixtures concerns the character of atom-atom interactions. While interactions within the nuclear spin-manifold are characterized by the same s-wave scattering length, leading to the SU($N$) symmetry discussed in Sec. \ref{sec:su(n)}, the scenario completely changes when atoms in different electronic states are considered: the scattering lengths now strongly depend on the internal state. Let's consider fermionic $^{173}$Yb as an example. The s-wave scattering lengths for binary $g-g$ and $e-e$ collisions are \cite{kitagawa2008,scazza2014}
\begin{align}
& a_{gg} \simeq +199 a_0 \\
& a_{ee} \simeq +310 a_0 \, ,\nonumber
\end{align}
where $a_0$ is the Bohr radius. Of course these scattering lengths refer to collisions between particles in different nuclear-spin states\footnote{For particles in the same electronic \textit{and} nuclear-spin state s-wave collisions are forbidden by the fermionic statistics.}. When collisions in different nuclear-spin \textit{and} electronic states are considered, two scattering lengths have to be considered, according to the exchange symmetry of the two-particle state. As a matter of fact, two identical fermions, each in a different electronic state $|g\rangle$ and $|e\rangle$ and in a different nuclear-spin state $|$$\uparrow$$\rangle$ and $|$$\downarrow$$\rangle$ \footnote{With $\uparrow$ and $\downarrow$ we indicate two generic spin projection states $m$ and $m'$ out of the nuclear-spin manifold.}, can be found in two different symmetrized two-particle states:
\begin{align}
& |eg^+\rangle = \frac{1}{2} \left( |g_1e_2\rangle + |e_1g_2\rangle \right)
\otimes
\left( |\uparrow_1\downarrow_2 \rangle - |\downarrow_1\uparrow_2 \rangle \right) = \frac{1}{\sqrt{2}} \left(
|g \uparrow, e\downarrow \rangle - |g \downarrow, e\uparrow  \rangle \right)
\\
& |eg^-\rangle =\frac{1}{2}  \left( |g_1e_2\rangle - |e_1g_2\rangle \right)
\otimes
\left( |\uparrow_1\downarrow_2 \rangle + |\downarrow_1\uparrow_2 \rangle \right)   = \frac{1}{\sqrt{2}} \left(
|g \uparrow, e\downarrow \rangle + |g \downarrow, e\uparrow  \rangle \right)
 \, ,\nonumber
\end{align}
where the subscript $\pm$ in the definition of the two-particle states $|eg^\pm\rangle$ refers to the exchange symmetry of the electronic state. As it can be seen from the last equivalence at the right-hand side (where the exchange symmetry is left implicit to ease the notations), in both $|eg^+\rangle$ and $|eg^-\rangle$ the nuclear-spin orientation for the atom in a given electronic state is not defined, being it in a superposition state (with entanglement between the two atoms). The scattering lengths associated to these two-particle states in $^{173}$Yb turn out to be quite different \cite{scazza2014,hofer2015,cappellini2019}:
\begin{align}
& a_{eg^+} \simeq +1890 a_0\\
& a_{eg^-} \simeq +220 a_0 \, .\nonumber
\end{align}
As a consequence, a strong spin-exchange process  can be observed when the atoms are prepared in an initial state with well defined nuclear-spin orientation. If we consider $|g \uparrow, e\downarrow \rangle$ as initial state, the time evolution induced by the interaction Hamiltonian
\begin{equation}
\hat{H}=U_{eg^+} |eg^+\rangle \langle eg^+| +
U_{eg^-} |eg^- \rangle \langle eg^-|
\, ,
\end{equation}
with $U_{eg^\pm}$ the interaction energy associated to each of the two-particle states (defined in the spirit of the Fermi-Hubbard model in Eq. (\ref{eq:fermihubbard})), leads to a time-evolved state
\begin{align}
|\Psi(t)\rangle & = e^{-\mathrm{i} t \hat{H}/\hbar} |g \uparrow, e\downarrow \rangle =\\
&= e^{-\mathrm{i} t \hat{H}/\hbar}  \frac{1}{\sqrt{2}}  \left( |eg^+\rangle + |eg^-\rangle \right) \nonumber \\
&=   \frac{1}{\sqrt{2}}  \left( e^{-\mathrm{i} t U_{eg^+}/\hbar} |eg^+\rangle + e^{-\mathrm{i} t U_{eg^-}/\hbar} |eg^-\rangle \right) 
=\nonumber \\
&=   \frac{ e^{-\mathrm{i} t U_{eg^+}/\hbar}+e^{-\mathrm{i} t U_{eg^-}/\hbar} } {2} |g \uparrow, e\downarrow \rangle \nonumber \, - \frac{ e^{-\mathrm{i} t U_{eg^+}/\hbar}-e^{-\mathrm{i} t U_{eg^-}/\hbar} } {2} |g \downarrow, e\uparrow \rangle \nonumber \,,
\end{align}
which exhibits a clear interorbital spin-exchange dynamics, where the spin state of the $g$ atom oscillates between $\uparrow$ and $\downarrow$ (and so that of the $e$ atom), with a probability for the two spin configurations
\begin{align}
\label{eq:spinexchange}
|\langle g \uparrow, e\downarrow | \Psi(t) \rangle |^2 & = \cos^2 \left( \frac{V_{ex}t}{\hbar}\right)  \\
|\langle g \downarrow, e\uparrow | \Psi(t) \rangle |^2 & = \sin^2 \left( \frac{V_{ex}t}{\hbar}\right) \nonumber \, ,
\end{align}
where we have defined $V_{ex}=(U_{eg^+}-U_{eg^-})/2$ as an interorbital local {\it exchange} energy. This spin-exchange dynamics for $^{173}$Yb fermions was first studied in Ref. \cite{scazza2014}, with the eventual observation of coherent spin oscillations in Ref. \cite{cappellini2014}. Fig. \ref{fig:spinexchange} reports the results of a Florence experiment where  $|g \uparrow, e\downarrow \rangle$ atom pairs in a deep 3D optical lattice were left free to evolve, resulting in an oscillation of the $g$ magnetization, which is a direct manifestation of the dynamics described in Eq. (\ref{eq:spinexchange})\footnote{The contrast of the oscillation is just a few percent because of technical limitations, mostly on the timescale for the magnetic-field switching that was used to initialize the spin dynamics.}.

This interorbital spin-exchange interaction is the key ingredient for the realization of a variety of Hamiltonians for the description of strongly correlated materials \cite{gorshkov2010}, most notably the Kondo lattice model, or advanced models supporting Majorana excitations \cite{iemini2017}. In addition to $^{173}$Yb, where this spin exchange  is ferromagnetic, recent experiments have studied the properties of $^{171}$Yb, where the interaction was found to be antiferromagnetic \cite{ono2019}, with interesting perspectives for the simulation of the Kondo effect.

\subsection{Orbital Feshbach resonance and orbital molecules}

The existence of two distinct interaction channels $|eg^+ \rangle$ and $|eg^- \rangle$, together with the possibility of mixing them by applying a magnetic field, also led to the prediction of the existence of an {\it orbital} Feshbach resonance in $^{173}$Yb \cite{zhang2015}, which was eventually observed in experiments in Florence and at MPQ \cite{pagano2015,hofer2015}. The mechanism for this Feshbach resonance, involving collisions of atoms in two electronic {\it and} nuclear-spin states $|g\uparrow\rangle$ and $|e\downarrow\rangle$, is similar to that of ordinary magnetic Feshbach resonances involving alkali atoms in different hyperfine states \cite{chin2010}. There are, however, distinct properties, most notably the fact that this resonance, located at a conveniently small magnetic field (just few tens of Gauss in free space, depending on the choice of nuclear-spin states), has a {\it narrow} character, but it is characterized by a {\it large} magnetic-field width\footnote{The reason for this apparent contradiction can be understood in the strongly suppressed magnetic-field sensitivity of the nuclear spin with respect to the nonzero electron spin of an alkali atom: the small width of the orbital Feshbach resonance in energy space (typically less than the Fermi energy of the system) is thus reflected in a strong range of magnetic fields (tens of Gauss rather than a few milliGauss), allowing for its convenient control.}. 

As an example of application of the orbital Feshbach resonance in $^{173}$Yb we mention the controlled creation and coherent manipulation of {\it orbital} molecules formed by two atoms in different electronic states \cite{cappellini2019}, with interesting perspectives both for metrological applications and for the investigation of strongly-interacting many-body states, potentially hosting unconventional superfluid states \cite{xu2016} (see Ref. \cite{zhang2020} for a review).


\acknowledgments
I thank M. Inguscio, J. Catani and G. Cappellini for the continuous and stimulating collaboration over the years, and all the students and postdocs who have worked on the Florence Yb experiment. Most of the experimental results presented in these lectures have been obtained within the ERC Consolidator Grant TOPSIM ``Topology and symmetries in synthetic fermionic systems" (Grant Agreement no. 682629), which is acknowledged for funding.


\end{document}